\documentclass[prb,twocolumn,showpacs,amsmath,amssymb,preprintnumbers,superscriptaddress]{revtex4}
\usepackage{dcolumn}
\usepackage{bm}
\usepackage{graphicx}
\usepackage{color}
\usepackage[utf8]{inputenc}
\DeclareUnicodeCharacter{2212}{-}

\begin{document}

\title{Structure, magnetism and electronic properties in 3$d$-5$d$ based double perovskite (Sr$_{1-x}$Ca$_x$)$_2$FeIrO$_6$ (0 $\leq$ $x$ $\leq$ 1): A combined experimental and theoretical investigation}

\author{K. C. Kharkwal}\affiliation{School of Physical Sciences, Jawaharlal Nehru University, New Delhi - 110067, India}
\author{Roumita Roy}\affiliation{School of Physical Sciences, Indian Institute of Technology Goa, Goa - 403401, India}
\author{Harish Kumar}\affiliation{School of Physical Sciences, Jawaharlal Nehru University, New Delhi - 110067, India}
\author{A. K. Bera}\affiliation{Solid State Physics Division, Bhabha Atomic Research Centre, Mumbai - 400085, India}
\author{S. M. Yusuf}\affiliation{Solid State Physics Division, Bhabha Atomic Research Centre, Mumbai - 400085, India}
\author{A. K. Shukla}\affiliation{National Physical Laboratory, Dr. K.S. Krishnan Road, New Delhi - 110012, India }
\author{Kranti Kumar}\affiliation{UGC-DAE Consortium for Scientific Research, Indore - 452001, India.}
\author{Sudipta Kanungo}\email{sudipta@iitgoa.ac.in}\affiliation{School of Physical Sciences, Indian Institute of Technology Goa, Goa - 403401, India}
\author{A. K. Pramanik}\email{akpramanik@mail.jnu.ac.in}\affiliation{School of Physical Sciences, Jawaharlal Nehru University, New Delhi - 110067, India}

\begin{abstract}
The 3$d$-5$d$ based double perovskites offer an ideal playground to study the interplay between electron correlation ($U$) and spin-orbit coupling (SOC) effect, showing exotic physics. The Sr$_2$FeIrO$_6$ is an interesting member in this family with ionic distribution of Fe$^{3+}$ (3$d^5$) and Ir$^{5+}$ (5$d^4$) where the later is believed to be nonmagnetic under the picture of strong SOC. Here, we report detailed investigation of structural, magnetic and electronic transport properties along with electronic structure calculations in (Sr$_{1-x}$Ca$_x$)$_2$FeIrO$_6$ series with $x$ from 0 to 1. While the basic interactions such as, $U$ and SOC are unlikely to be modified but a structural modification is expected due to ionic size difference between Sr$^{2+}$ and Ca$^{2+}$ which would influence other properties such as crystal field effect and band widths. While a nonmonotonic changes in lattice parameters are observed across the series, the spectroscopic investigations reveal that 3+/5+ charge state of Fe/Ir continue till end of the series. An analysis of magnetic data suggests similar nonmonotonic evolution of magnetic parameters with doping. Temperature dependent crystal structure as well as low temperature (5 K) magnetic structure have been determined from neutron powder diffraction measurements which further indicate a site ordered moments for both Fe and Ir. The whole series shows insulating behavior with a nonmonotic variation in resistivity where the charge transport follows the 3-dimensional variable range hopping model. The electronic structure calculations show, SOC enhanced, a noncollinear antiferromagnetic and Mott-type insulating state is the stable ground state for present series with a substantial amount of orbital moment, but less than the spin magnetic moment, at the Ir site and the magnetocrystalline anisotropy. The calculations further show the evolution of the spin and orbital magnetic moment components across the series along with the magnetization density. The obtained results imply that the local structural modification with introduction of lower size Ca$^{2+}$ has large influence on the magnetic and transport properties, further showing a large agreement between experimental results as well as theoretical calculations.  
\end{abstract}


\maketitle

\section{Introduction}
An alternatively arranged, two interpenetrating face centered cubic (FCC) sublattices of $B$ and $B'$ transition metal cations, govern most of the physical properties in double perovskites (DPs) with chemical composition $A_2BB'$O$_6$ where $A$ is alkaline or rare earth element.\cite{Longo,Sleight,Yokoyama} The $B$ and $B'$ are generally octahedrally coordinated with the oxygen anions. The magnetic character of transition metal cations plays an important role as it introduces geometrical frustration that arises due to corner shared tetrahedra of either $B$ or $B'$ in the FCC sublattices. The sizes of the $A$, $B$ and $B'$ cations further decide the structural stability of the DPs.\cite{Serrate, Vasala} Mismatch in the sizes of cations distort the crystal structure to lower symmetry and this mismatch is conventionally measured using tolerance factor ($t$),

\begin{equation}
t = \frac{r_A +r_O}{\sqrt{2}((r_B +r_{B'})/2 + r_O)}
\end{equation}

where $r_i$ is the respective ionic radii of $A$, $B$, $B'$ and O ions.\cite{Serrate,Vasala} With lowering of $t$, the most commonly occurred tilt systems in DPs are $Fm-3m$ $\rightarrow$ $I4/m$ $\rightarrow$ $I2/m$ $\rightarrow$ $P2_1/m$ $\rightarrow$ $I\bar{1}$. The tilt in structural symmetry leads to modification in magnetic and transport properties in these materials accordingly.\cite{Serrate,Vasala}

In DPs, there are several choices for $A$-site atom starting from alkaline- to rare-earth element, having different valance states and ionic sizes.\cite{Vasala} Changes in valence state or ionic size of $A$-site cation play crucial role in tuning of magnetic as well as electronic properties.\cite{Vasala} For instance, substitution of smaller size Ca$^{+2}$ for Sr$^{2+}$ introduces local structural distortion which even distorts the crystal structure to lower symmetry, where the various magnetic ground states with different electronic properties are realized. Prominent example is Sr$_2$FeReO$_6$ where introduction of Ca$^{2+}$ not only drives the system from metallic to insulating state,\cite{Gopal} but also diverse magnetic states such as, spin glass (SG) or antiferromagnetic (AFM) have been found.\cite{Battle,Nomura,Naveen} Previous studies have further shown a tetragonal structure with AFM ground state in Sr$_2$FeOsO$_6$ whereas Ca$_2$FeOsO$_6$ shows a distorted monoclinic structure and ferromagnetic (FM) state with a high transition temperature.\cite{Paul,Feng,Kanungo1} Interesting properties have also been observed in materials other than DPs. For instance, a suppression of FM state and resultant `quantum phase transition' have been observed in perovskite Sr$_{1-x}$Ca$_x$RuO$_3$.\cite{Cao1,Gat,Fuchs} Further interesting results include spin configuration change in (Sr$_{1-x}$Ca$_x$)$_3$YCo$_4$O$_{10.5}$,\cite{Terasaki} redistribution of charge along chains in Sr$_{14-x}$Ca$_x$Cu$_{24}$O$_{41}$, suppression of anisotropic magnetoresistance in La$_{0.67}$(Ca$_{1-x}$Sr$_x$)$_{0.33}$MnO$_3$,\cite{Ma,Deng,Liu} crossover from FM and Fermi-liquid to AFM and metallic state in Sr$_{1-x}$Ca$_x$Co$_2$P$_2$,\cite{Jia} etc.  

The 3$d$-5$d$ based DPs are of recent interest where many exotic states of matter may emerge due to fine interplay between electronic correlation ($U$) and spin-orbit coupling (SOC) which are prominent in 3$d$ and 5$d$ elements, respectively.\cite{Philipp,Taylor,Chen,Krock,Krempa,Rau} Iridium is one of the important choice for 5$d$ element because of its strong SOC and complex oxidation states (4+/5+).\cite{Kim, Kim1,Kharkwal,Wallace,Nag} Within the oxygen environment, the $d$ orbitals of Ir split into $e_g$ and $t_{2g}$ states where the $t_{2g}$ state under the influence of strong SOC further splits into low lying $J_{eff}$ = 3/2 quartet and $J_{eff}$ = 1/2 doublet. Following this model, a full-filled $J_{eff}$ = 3/2 and a half-filled $J_{eff}$ = 1/2 state is realized for Ir$^{4+}$ (5$d^5$). The $J_{eff}$ = 1/2 band, even in presence of small $U$, opens up a gap around 0.5 eV, thus representing an exotic example of $J_{eff}$ = 1/2 Mott insulator.\cite{Kim, Kim1} The Ir$^{5+}$ (5$d^4$), on the other hand, has an empty $J_{eff}$ = 1/2 state, therefore it is considered to have nonmagnetic ground state ($J_{eff}$ = 0).\cite{Khaliullin} However, the emergence of SOC driven $J_{eff}$ states is under debate. The breakdown of $J_{eff}$ picture in iridates under strong noncubic crystal field in the distorted IrO$_6$ octahedra has been reported for both Ir$^{4+}$ and Ir$^{5+}$ systems. For example, the Ir$^{4+}$ systems viz. CaIrO$_3$, Sr$_2$CeIrO$_6$ are the examples of breakdown of strong $J_{eff}$=1/2 picture.\cite{Sala,Kanungo2} On the other side, the breakdown of $J_{eff}$ = 0 state in Ir$^{5+}$ has been evidenced by emergence of non-zero magnetic moment in prototype systems of Sr$_2$YIrO$_6$, Ba$_2$YIrO$_6$, Ba$_3$ZnIr$_2$O$_9$.\cite{Cao,Ranjbar,Dey,Phelan,Nag} An excitation from $J$ = 0 (singlet) to low lying $J$ = 1 (triplet) and 2 (quintet) states are expected to provide exotic magnetism in $d^4$ systems theoretically, however, recent moment dependent resonant inelastic x-ray scattering (RIXS) spectra analysis reveal that these low lying states are unlikely since the exchange strength between Ir-Ir is not sufficiently high to overcome the magnetic gap.\cite{Kusch,Kim2} These suggest the origin of magnetism in iridium is still not very clear and needs further investigations in other similar materials.

In the present work, we have studied detailed structural, magnetic and electronic transport properties of 3$d$-5$d$ based DPs (Sr$_{1-x}$Ca$_x$)$_2$FeIrO$_6$ (0 $\leq$ $x$ $\leq$ 1). We have further done band structure calculations to understand the evolution of magnetic and electronic structure in present series. The ionic sizes of Sr$^{2+}$ and Ca$^{+2}$ are 1.44 {\AA} and 1.34 {\AA} (12 coordination number), and those of Fe$^{3+}$ (high spin state) and Ir$^{5+}$ are 0.645 {\AA} and 0.57 {\AA} (6 coordination number), respectively.\cite{Shannon} These reasonable differences between the ionic radii and charge states of transition metals would minimize the anti-site disorder (i.e., occupation of Fe site by Ir and vice-versa). The substitution of isoelectronic Ca$^{2+}$ for Sr$^{2+}$ will not tune the vital parameters such as, SOC and $U$, but is expected to introduce (local) structural distortion through modification of bond-length and bond-angle between transition metals and oxygen. Therefore, the environment of individual FeO$_6$/IrO$_6$ octahedra will be modified which would affect the three dimensional network of Fe-O-Ir-O-Fe. The introduced structural distortion with modified bandwidth will eventually compete with the other electronic energy scales (such as, SOC and $U$), and based on its strength would introduce anisotropy, orbital magnetism, spin canting or even spin state transition in transition metals, etc. These will have large ramification on the magnetic and transport properties given that transition metal oxides exhibit prominent interrelation between lattice, spin, charge and orbital degrees of freedom. The results will further help to understand whether conventionally accepted $J_{eff}$ = 0 state for Ir$^{5+}$ (5$d^4$) still remain valid in distorted crystal structure. Therefore, our interest is to investigate how the properties evolve from one end ($x$ = 0) to the other end ($x$ = 1) in present series while the driving parameter is the structural distortion introduced by an ionic size mismatch. Further, the whole series ($x$ = 0 to 1) has been studied to examine any nonmonotonic evolution of parameters, as observed in other similar series.\cite{Cao1,Fuchs,Jia}

The Sr$_2$FeIrO$_6$ ($x$ = 0) shows a triclinic ($I\bar{1}$) or monoclinc ($P2_1/n$, $I2/m$) crystal structure and a long-range AFM-type transition around 120 K.\cite{Battle1,Qasim,Bufaical,Kayser,Laghuna} In our previous study,\cite{Kharkwal} we have though shown two magnetic transitions in Sr$_2$FeIrO$_6$; one prominent transition at $\sim$ 45 K and other weak one at $\sim$ 120 K. The similar dual AFM transitions with different spin structure are observed in other 3$d$-5$d$ based DP Sr$_2$FeOsO$_6$.\cite{Avijit,Morrow} Another end compound Ca$_2$FeIrO$_6$ ($x$ = 1) has been shown to have monoclinc ($P2_1/n$) structure and a low temperature AFM transition at $\sim$ 75 K.\cite{Bufaical} Our structural investigation reveal that while the original crystal symmetry is retained across the series but the lattice parameters modify over the series. Spectroscopic analysis indicate a 3+/5+ charge state for Fe/Ir is retained over the series. In fact, we find that the structural, magnetic and electronic parameters exhibit not only a nonmontonic evolution with Ca ($x$) but their variations are correlated. The band structure calculations predict an AFM-insulating phase is the stable ground state in these materials under the combined influence of $U$ and SOC. Further, an evolution of magnetic moment component ($m_x$, $m_y$, $m_z$) with Ca concentration is predicted which mainly arises due to local structural distortion. Importantly, both the neutron measurements and the calculation show an agreement for the moment of Ir,

\section{Experimental Details}
Polycrystalline samples of (Sr$_{1-x}$Ca$_x$)$_2$FeIrO$_6$ ($x$  = 0.0, 0.05, 0.1, 0.2, 0.4, 0.6, 0.8, 1.0) series are prepared using solid state reaction method by calcination and sintering in high temperatures. The high purity ingredient powders (M/s Sigma Aldrich) of SrCO$_3$, CaCO$_3$, Fe$_2$O$_3$ and IrO$_2$ are taken in stoichiometric ratio and ground well. The ground fine powders are calcined in air at 900$^{\circ}$ for 24 h twice with an intermediate grinding. Calcined powders are then pressed into pallets and sintered in air at 1000$^{\circ}$C, 1050$^{\circ}$C and 1100$^{\circ}$C for 24 h each time with intermediate grindings. A horizontal tube furnace has been used for high temperature treatment. Given that IrO$_2$ is volatile in high temperature, a relatively low temperature required for present material synthesis minimizes the issue of IrO$_2$ volatilization. Iridium loss has though not been accounted in calculation. As these materials are synthesized in relatively low temperatures, a small amount of porosity may be present in the materials. The synthesis temperature, however, could not be raised further to avoid any serious loss of Ir. All the materials are prepared in identical condition which would give similar material quality in terms of Ir/Fe content, stiochiometry as well as porosity. Further, these materials are highly insulating, therefore a small amount of porosity would not affect the overall conductivity severely. The chemical phase purity of the samples are primarily checked with powder x-ray diffraction (XRD) where the data have been collected using Rigaku MiniFlex diffractometer equipped with Cu-K$_{\alpha}$ source. The XRD data have been analyzed with Rietveld refinement program (FullProf software).\cite{Full}

To probe the charge state of constituent elements, x-ray photoemission spectroscopy (XPS) measurements have been done. The XPS measurements have been performed at base pressure of 10$^{-11}$ mbar with an Omicron ultra-high vacuum (UHV) chamber equipped with a non-monochromatic Mg-K$_{\alpha}$ x-ray source ($h\nu$ = 1253.6 eV) and a multi-channeltron hemispherical electron energy analyzer (EA 125). The XPS data have been analyzed using XPS peak-fit 4.1 software. DC magnetization data are collected using physical property measurement system (Quantum  Design). The temperature dependent magnetization data have been collected in sweep mode with a temperature sweeping rate 1.5 K/min. Neutron powder diffraction (NPD) pattern over a wide angular range (3-140$^{\circ}$) are recorded at room temperature (counting time $\sim$ 6 hrs) using the powder diffractometer PD-2 ($\lambda$ = 1.2443 {\AA}) at Dhruva reactor, Trombay, India. Additional neutron diffraction patterns have been recorded at 5 and 150 K (counting time $\sim$ 30 hrs per temperature), to measure the magnetic Bragg peaks using the limited angular-range powder diffractometer PD-1 ($\lambda$  = 1.094 {\AA}) at  Dhruva reactor, Trombay, India. For the NPD measurements, powder samples ($\sim$ 4 g) are filled in a vanadium can of diameter around 6 mm where the low temperature measurements have been done using closed cycle refrigerator (CCR). The neutron flux during these experiments was $\sim$ 5 $\times$ 10$^5$ neutrons/cm$^2$/sec. The NPD patterns are analyzed by Rietveld refinement program.\cite{Full} Temperature dependent electrical conductivity data are collected in a CCR based home-made insert and the magnetic field dependent electrical conductivity have been measured in an insert attached with 9 Tesla magnet (Nanomagnetics). The conductivity has been measured using the standard four-probe method with sintered pressed pellets for all the compositions. The data are collected in a sweep mode with temperature sweeping rate around 1.5 K/min. 

\section{Computational Details}
Density functional theory (DFT) based electronic structure calculations are performed using the plane-wave basis set with a pseudopotential framework as incorporated in the Vienna ab-initio simulation package (VASP).\cite{vasp} The generalized gradient  exchange-correlation approximated (GGA) functional was employed following the Perdew–Burke–Ernzerhof (PBE) prescription.\cite{PBE} The missing on-site electron-electron Coulomb correlation are taken through the inclusion Hubbard ($U$) in GGA+$U$.\cite{GGAU1,GGAU2} The spin-orbit coupling (SOC) is introduced as a scalar relativistic correction to the original Hamiltonian. The plane-wave cut-off is employed to be 500 eV. A $k$-point mesh of $8 \times 8 \times 6$ in the Brillouin zone (BZ) has used for self-consistent calculations. The coordinates of the atomic positions are relaxed toward equilibrium until the Hellmann-Feynman force becomes less than 0.001 eV/\AA keeping the lattice parameters fixed at the experimentally determined values.

\section{Results and Discussion}
\subsection{Structural analysis using x-ray diffraction}
Figure 1 shows the XRD patterns for (Sr$_{1-x}$Ca$_x$)$_2$FeIrO$_6$ series at room temperature. As evident in figure, the
symmetry in the structural data is retained over the series. However, with progressive substitution of Ca, the major observations are; intensity of peak at 2$\theta$ = 23.2$^{\circ}$ increases, a small splitting arises in peak situated at 2$\theta$ = 59.1$^{\circ}$ and the position of peaks shifts to higher 2$\theta$ value. Given that the ionic radii of Ca$^{2+}$ (1.34 \AA) and Sr$^{2+}$ (1.44 \AA) has some differences, therefore even if a structural phase transition does not occur but a structural modification/distortion at local level is expected with the substitution of Ca for Sr.

\begin{figure}
\centering
		\includegraphics[width=8cm]{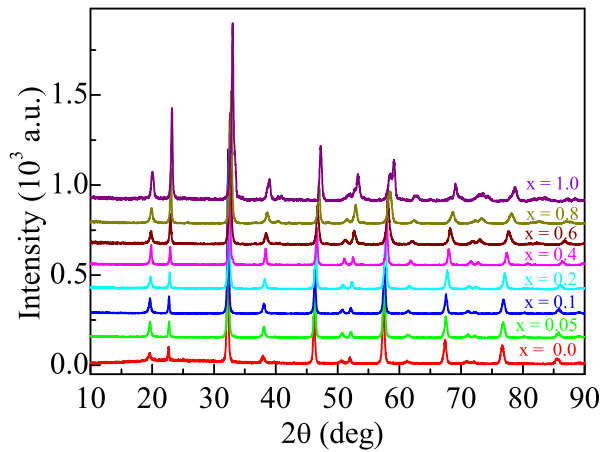}
\caption{(color online) XRD data collected at room temperature are shown for (Sr$_{1-x}$Ca$_x$)$_2$FeIrO$_6$ series with $x$ = 0.0, 0.05, 0.1, 0.2, 0.4, 0.6, 0.8, 1.0.}
	\label{fig:Fig1}
\end{figure}

Representative XRD pattern along with Rietveld refinement have been shown for two end members i.e., $x$ = 0.0 and 1.0 of present series in Fig. 2. The crystal structure of Sr$_2$FeIrO$_6$ ($x$ = 0.0) has already been reported by several groups including us. All the reports show a distorted structure but debate is between monoclinic and triclinic structure. For instance, Battle \textit{et al.}\cite{Battle1} has shown a monoclinic-$P2_1/n$ structure from XRD data but their high-resolution neutron powder diffraction data indicate a triclinic-$I\bar{1}$ structure. Similarly, other groups have shown monoclinic-$P2_1/n$ or monoclinic-$I2/m$ structure using laboratory XRD, synchrotron XRD and powder neutron diffraction data.\cite{Qasim,Bufaical} We have tried Rietveld refinement with all the above mentioned crystal structures for room temperature XRD data but out results show best fitting for triclinic-$I\bar{1}$ crystal structure as gauged by lowest $\chi^2$ value. For example, we obtain $\chi^2$ values for triclinic-$I\bar{1}$, monoclinic-$P2_1/n$ or monoclinic-$I2/m$ structures are 4.78, 5.28, 5.17, respectively.\cite{Kharkwal} We also obtain similar lowest $\chi^2$ value for triclinic-$I\bar{1}$ structure in Rietveld analysis with neutron powder diffraction data (discussed later). We have further verified, by DFT total energy calculations, taking into account both electron correlation and SOC effect for both triclinic and monoclinic symmetry. The results indicate that although the energetics are very similar, however, the triclinic symmetry structure is energetically lower than the monoclinic symmetry structure by an amount of $\sim$ 2 meV/f.u.

Figure 2 in the present manuscript shows reasonably good Rietveld fitting with triclinic structure. Further, on close inspection, the XRD data in Figs. 1 and 2 show multiple peaks related to distorted structure which is not though very clear in figures because XRD has relatively low resolution. However, neutron powder diffraction pattern in Fig. 9a indeed show clear multiple peaks, and all the peaks are refined by Rietveld analysis with triclinic-$I\bar{1}$ structural symmetry. Here, it can be mentioned that the results of all Rietveld analysis and energy calculations show a small difference between triclinic and monoclinic structure, but the triclinic structure consistently shows a stable structural phase for present materials. Fig. 2a displays the Rietveld refinement of XRD data with triclinic-\textit{I$\bar{1}$} symmetry for Sr$_2$FeIrO$_6$ ($x$ = 0.0) sample, showing a reasonably good fitting. In fact, we find that the whole series in the present study can be fitted well with the same triclinic-\textit{I$\bar{1}$} crystal symmetry, as indicated by lowest $\chi^2$ value. Fig. 2b shows the same Rietveld fitting for Ca$_2$FeIrO$_6$ ($x$ = 1.0) sample. The $\chi^2$ values obtained from Rietveld refinement are 4.78, 2.62, 3.07, 2.20, 3.18, 2.03, 2.28, 2.71 for $x$ = 0.0, 0.05, 0.1, 0.2, 0.4, 0.6, 0.8, 1.0 composition, respectively which signifies a good fitting of XRD data. Here, it can be mentioned that using the real structural parameters from XRD analysis, we calculate the tolerance factor ($t$) to be 0.95 and 0.77 for $x$ = 0.0 and 1.0, respectively which indicates a distorted crystal structure (monoclinic or triclinic) for these materials. Point to be noted here that, most of the Ir based double perovskites (Sr,La)$_2$$M$IrO$_6$ ($M$ = Co, Mg, Zn, Cu, Ce) are reported to adopt distorted monoclinic or triclinic structure, as found in the literature.\cite{Narayanan,GCao,WKZhu,Kanungo2}
 
\begin{figure}
\centering
		\includegraphics[width=8cm]{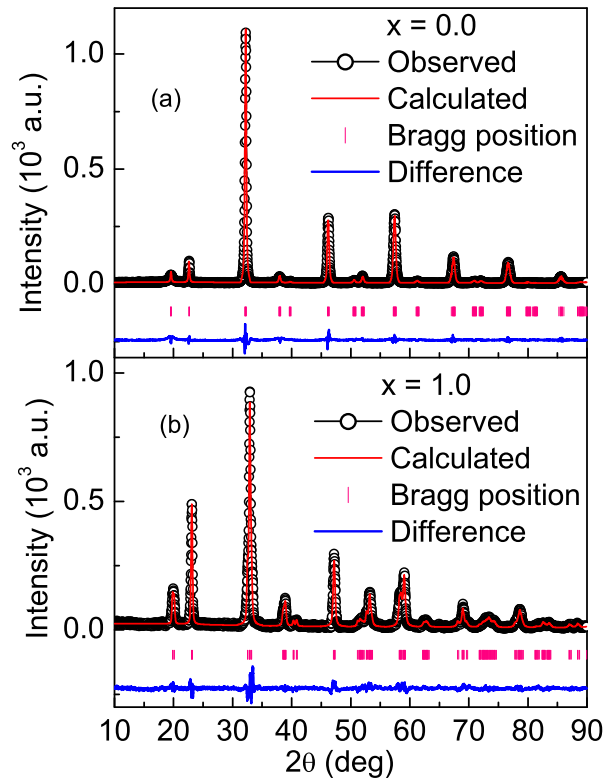}
\caption{(color online) (a) shows the Rietveld refinement of powder x-ray diffraction data with triclinic-\textit{I$\bar{1}$} symmetry for $x$ = 0.0 compound in (Sr$_{1-x}$Ca$_x$)$_2$FeIrO$_6$ series. The experimentally observed intensity (open circle), calculated intensity (red line) and their difference (blue line) are shown in figure. (b) shows the same for $x$ = 1.0 compound of present series.}
	\label{fig:Fig2}
\end{figure}

The unit cell parameters ($a$, $b$, $c$, $\alpha$, $\beta$ and $\gamma$), extracted from the refinement of XRD data are shown in Fig. 3 as a function of doping concentration $x$. Although the triclinic-\textit{I$\bar{1}$} symmetry continues for whole (Sr$_{1-x}$Ca$_x$)$_2$FeIrO$_6$ series, the crystal structure becomes more distorted with doping. Lattice constants $a$, $b$ and $c$ mostly decrease with doping. However, as seen in Fig. 3 all the structural parameters ($a$, $b$, $c$ and angles $\alpha$, $\beta$ and $\gamma$) show an anomalous variation around $x$ = 0.05 - 0.1 and then 0.6. The decrease in lattice constants $a$, $b$ and $c$ with doping concentration $x$ is due to smaller ionic radii of Ca$^{2+}$ compared to Sr$^{2+}$. Considering that this is an isoelectronic substitution, the doped Ca induces a chemical pressure at $A$-site to modify the lattice parameters.

\begin{figure}
\centering
		\includegraphics[width=8cm]{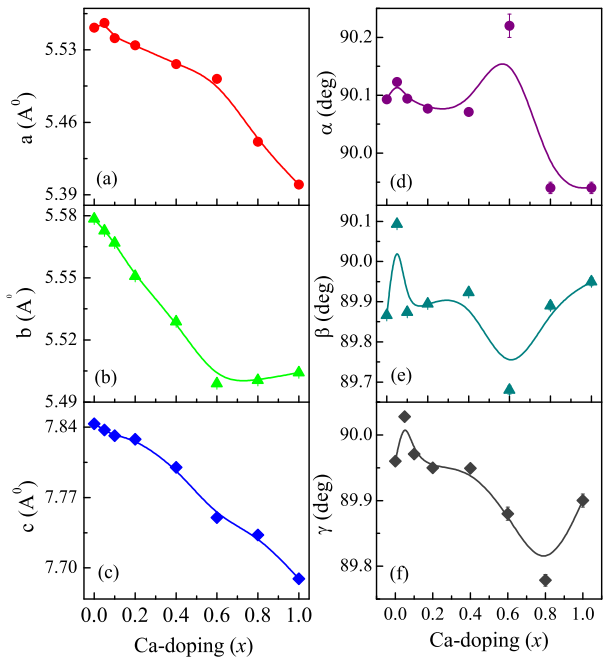}
\caption{(color online) Lattice parameters (a) $a$, (b) $b$, (c) $c$, (d) $\alpha$, (e) $\beta$, (f) $\gamma$ are plotted as a function of doping concentration $x$ for (Sr$_{1-x}$Ca$_x$)$_2$FeIrO$_6$ series. Lines are guide to eyes.}
	\label{fig:Fig3}
\end{figure}

\subsection {X-ray photoemission spectroscopy data}
\begin{figure*}
\centering
		\includegraphics[width=18cm]{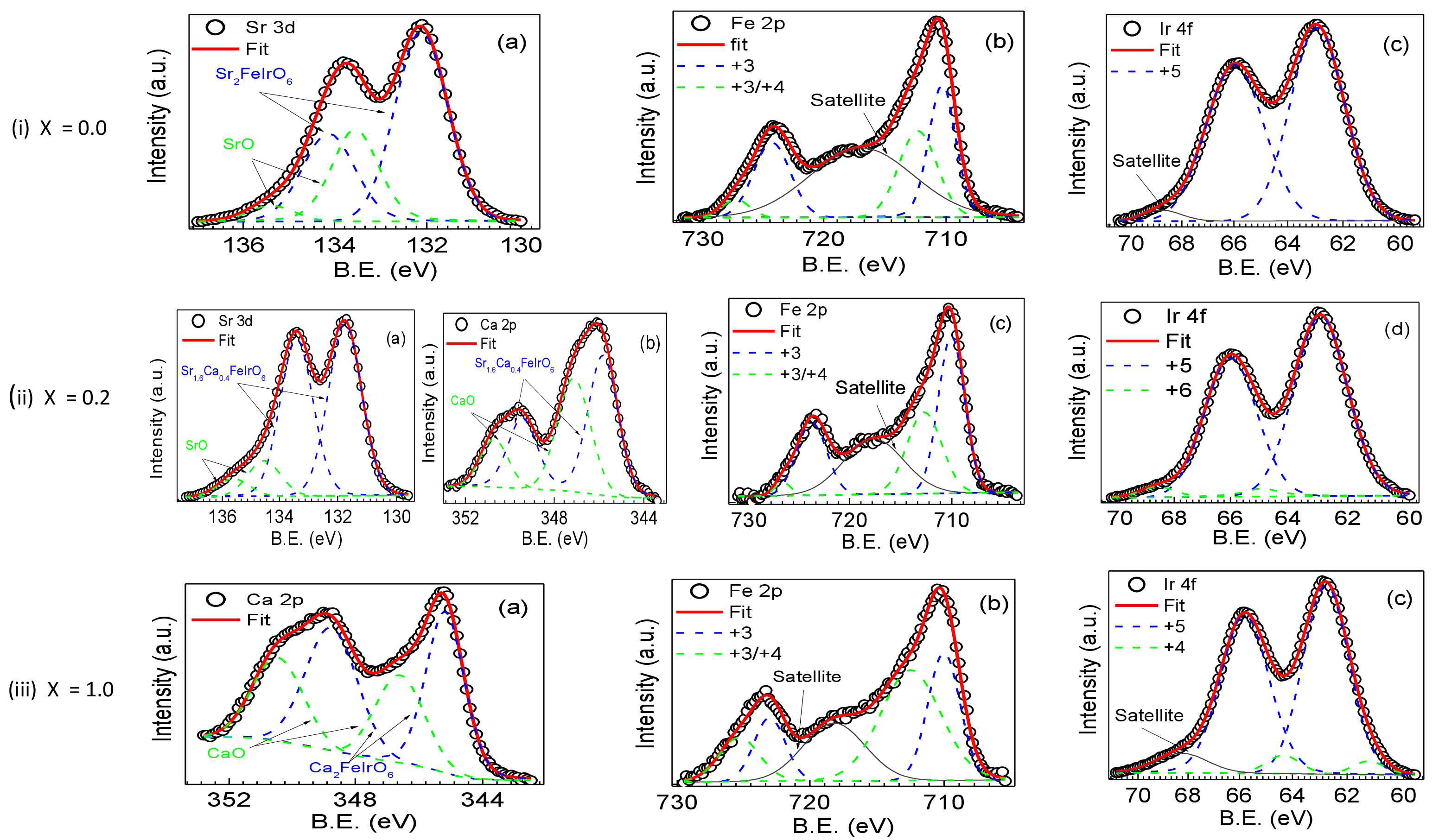}
\caption{(color online) The XPS spectrum are shown for two end members ($x$ = 0.0 and 1.0) and an intermediate compound with $x$ = 0.2 of (Sr$_{1-x}$Ca$_x$)$_2$FeIrO$_6$ series. Description of plots are; (i) $x$ = 0.0 [(a) Sr 3d, (b) Fe 2p and (c) Ir 4f ],(ii) $x$ = 0.2 [(a) Sr 3d, (b) Ca 2p, (c) Fe 2p and (d) Ir 4f ] and (iii) $x$ = 1.0 [(a) Ca 2p, (b) Fe 2p and (c) Ir 4f]. Red and gray solid lines are due to fitting of the whole spectrum and satellite, respectively whereas the individual fittings are shown in dotted blue and green lines (see text).}
	\label{fig:Fig4}
\end{figure*}

In order to understand the charge states of constituent transition metals i.e., Fe/Ir and its effect on physical properties in (Sr$_{1-x}$Ca$_x$)$_2$FeIrO$_6$ series, the XPS measurements have been performed for doping concentrations $x$ = 0.0, 0.2 and 1.0. The XPS data for representative $x$ = 0.0, 0.2 and 1.0 are shown in Figs. 4(i), 4(ii) and 4(iii), respectively as a function of binding energy (B.E). The Core levels have been fitted using Lorentzian components with small admixture of Gaussian. Figs. 4(i)a and 4(ii)a show Sr-3$d$ data for $x$ = 0.0 and 0.2, respectively, where the spin-orbit split Sr-3$d_{5/2}$ and Sr-3$d_{3/2}$ peaks are seen at B.E 132.1 and 133.9 eV, respectively (marked with dotted blue line). In addition, our fitting indicates similar Sr-3$d_{5/2}$ and Sr-3$d_{3/2}$ contribution from SrO (green dotted line) while the red solid lines in figures indicate the total fitting of the data.

The XPS data of Ca-2$p$ are shown for $x$ = 0.2 and 1.0 in Figs. 4(ii)b and 4(iii)a, respectively. The similar spin-orbit split Ca-2$p_{3/2}$ and Ca-2$p_{1/2}$ peaks of DP (Sr$_{1.6}$Ca$_{0.4}$FeIrO$_6$, Ca$_2$FeIrO$_6$,) and CaO are marked with dotted blue and green colors, respectively. The contribution from SrO/CaO in XPS spectra arises due to exposure of material surface to the atmosphere where oxidation of Sr/Ca takes place.\cite{Celorrio,Doveren,Liu1} The analysis of Sr-3$d$ and Ca-2$p$ spectra indicates that Sr/Ca is in 2+ charge state, as expected.

The XPS spectrum of Fe-2$p$ are shown in Figs. 4(i)b, 4(ii)c and 4(iii)b for $x$ = 0.0, 0.2 and 1.0, respectively. A pronounced satellite peak is seen in figure at about 8 eV above from the main peak. In XPS Fe-2$p$ spectrum, spin-orbit split 2$p_{3/2}$ and 2$p_{1/2}$ peaks are seen around 710 and 723.5 eV for Fe$^{3+}$ and around 712 and 725.5 eV for Fe$^{4+}$, respectively. Moreover, a pronounced satellite peak on top of shoulder is observed around 718 eV in case of Fe$^{3+}$ while a weak satellite feature is present in case of Fe$^{4+}$.\cite{Rogge,Ghaffari,Tsuyama} In view of these, the Fe-2$p$ XPS spectra shown in Figs. 4 for present materials suggest iron is in Fe$^{3+}$ state. Individual peaks related to Fe-2$p_{3/2}$ and Fe-2$p_{1/2}$ states for Fe$^{3+}$ charge state are marked in dotted blue while the satellite peak is shown in gray color. Using Mossbauer spectroscopy, Fe$^{3+}$ ionic state has been evidenced in Sr$_2$FeIrO$_6$.\cite{Battle} Previous studies on iron based materials report multiple peaks corresponding to Fe$^{3+}$ state where a satellite peak is seen exactly at 8 eV above from the main peak.\cite{Yamashita,Grosvenor,Mullet,Biesingera} The peaks shown with dotted green color in Fe-2$p$ spectra may be due to Fe$^{3+}$ complex, however, this also has been shown related to Fe$^{4+}$ state.\cite{Fuentes}

Figures 4(i)c, 4(ii)d and 4(iii)c show XPS data of Ir-4$f$ for $x$ = 0.0, 0.2 and 1.0 material, respectively. The Ir-4$f$ core level spectra are fitted well with two spin-orbit split peaks corresponding to Ir-4$f_{7/2}$ and Ir-4$f_{5/2}$ levels. In XPS Ir-4$f$ spectra, the spin-orbit split Ir-4$f_{7/2}$ and Ir-4$f_{5/2}$ doublets occur around 62 and 65 eV for Ir$^{4+}$ while for Ir$^{5+}$ those peaks occur at little bit higher energies, between 63-64 eV to 66-68 eV, respectively.\cite{Zhu,Imtiaz,Harish,Otsubo,Nag,Liu1} For $x$ = 0.0, these two peaks are observed at 62.9 and 65.9 eV with a satellite at 68.8 eV, where the difference between peaks is $\sim$ 3 eV. These peak positions are consistent with Ir$^{5+}$. The 3+/5+ charge state of Fe/Ir as obtained from the analysis of XPS data for Sr$_2$FeIrO$_6$ are consistent with the charge state obtained from the analysis of our magnetization data as well as with the electronic structure calculations (discussed later). For $x$ = 0.2, the peaks corresponding to Ir$^{5+}$ are observed at 62.9 and 66.0 eV along with a weak contribution of Ir$^{6+}$ state, seen at 64.9 and 68.6 eV. The peaks related to Ir$^{5+}$ and Ir$^{6+}$ are shown in dotted blue and green colors, respectively. Analysis of data reveals the content of Ir$^{5+}$ and Ir$^{6+}$ are about 95.1 and 4.9 \%, respectively. For $x$ = 1.0 material, on the other hand, we find a small trace of Ir$^{4+}$ along with Ir$^{5+}$ and a gray colored satellite peak. The peak positions of Ir$^{5+}$ are around 62.9 and 65.9 eV in the spectrum whereas those of Ir$^{+4}$ are found at 61.5 and 64.5 eV. The amount of Ir$^{5+}$ and Ir$^{4+}$ are analyzed to be around 94.1 and 5.9 \%, respectively. From the analysis of XPS data it is evident that iridium is mostly in Ir$^{5+}$ charge state in the present series while a small amount of neighboring charge states (i.e., Ir$^{4+}$ and Ir$^{6+}$) are observed mostly due to non-stoichiometry in materials.

\subsection{Magnetic study in (Sr$_{1-x}$Ca$_x$)$_2$FeIrO$_6$}
Our previous study has shown that the Sr$_2$FeIrO$_6$ has prominent long-range AFM transition around 45 K, however, a close observation in magnetization data reveals a weak AFM transition at higher temperature around 120 K which is marked by an onset of bifurcation between zero field cooled (ZFC) and field cooled (FC) magnetization data.\cite{Kharkwal} Similar magnetic transition around 120 K, though with different features, has been reported for Sr$_2$FeIrO$_6$ in previous studies.\cite{Battle1,Qasim,Bufaical} Further, an analysis of magnetization data for Sr$_2$FeIrO$_6$ gives an effective magnetic moment ($\mu_{eff}$) 6.19 $\mu_{B}$/f.u. which closely matches with the calculated value 5.92 $\mu_{B}$/f.u following the relation $\mu_{eff}$ = $g\sqrt{S(S+1)}$ where $S$ is the total spin, that indicates 3+/5+ charge state of Fe/Ir respectively.\cite{Kharkwal} Our XPS data analysis in Fig. 4 regarding the charge states of transition metals are in fact consistent with the magnetization data for Sr$_2$FeIrO$_6$.\cite{Battle1,Qasim,Bufaical,Kayser,Laghuna}
 
\begin{figure}
\centering
		\includegraphics[width=8cm]{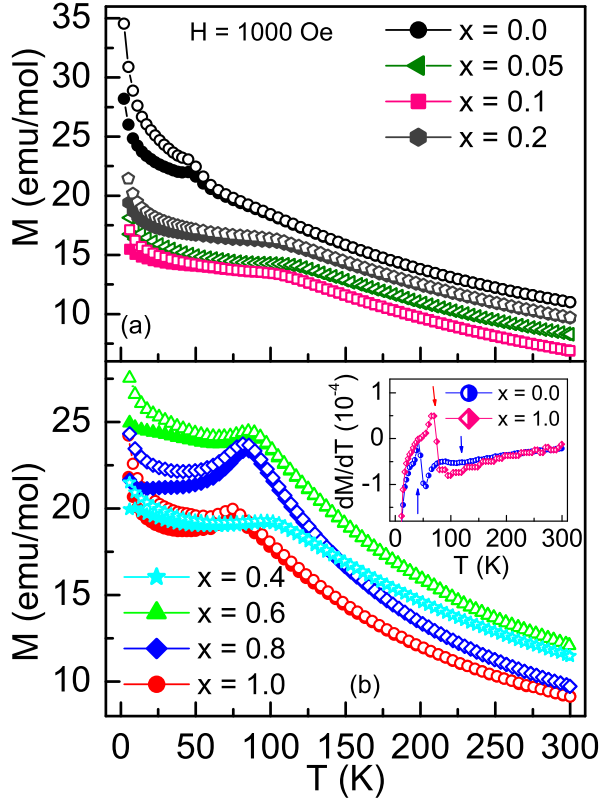}
\caption{(color online) DC magnetization data collected following ZFC and FC protocol at 1000 Oe magnetic field are shown as a function of temperature in (a) $x$ = 0.0, 0.05, 0.1, 0.2 and (b) $x$ = 0.4, 0.6, 0.8, 1.0 for (Sr$_{1-x}$Ca$_x$)$_2$FeIrO$_6$ series. The ZFC and FC data are represented in filled and open symbols, respectively. Inset of (b) shows the d$M$/dT plot for $x$ = 0.0 and $x$ = 1.0, where arrows indicate the transition temperatures of these materials.}
	\label{fig:Fig5}
\end{figure}

The ZFC and FC magnetization data collected in 1000 Oe magnetic field are shown in Figs. 5a and 5b for present (Sr$_{1-x}$Ca$_x$)$_2$FeIrO$_6$ series. While the nature of bifurcation between ZFC and FC magnetization data as well as the magnetic transition at T$_N$ changes with the Ca doping, the $M(T)$ shows a continuous decrease in high temperature PM
state for all the samples. In inset of Fig. 5b, temperature derivative of magnetization data (d$M$/dT) are shown for $x$ = 0.0 and 1.0 samples. As evident in figure, the d$M$/dT indicates a dual magnetic transition for Sr$_2$FeIrO$_6$  around 45 and 120 K, while for Ca$_2$FeIrO$_6$ a single transition is observed around 75 K which is in agreement with previous report.\cite{Bufaical} Further, the magnetic data show a relatively large moment in PM state and a small absolute value of d$M$/dT. While these could be due to an inhomogeneous magnetic state in PM state but the obtained moment agrees well with the previous reports of same material,\cite{Battle} and also with other Ir based paramagnetic material SrIrO$_3$ single crystals.\cite{Cao2} The composition dependent transition temperature T$_N$ is shown in Fig. 8a which shows T$_N$ decreases almost monotonically with $x$. Note that unlike Sr$_2$FeIrO$_6$, we have observed only single transition in doped materials, at least it is not clearly evident in magnetization data. We further mention that other 3$d$-5$d$ DP material i.e., Sr$_2$FeOsO$_6$ shows two AFM phase transitions ($T_N$ $\sim$ 140 and 67 K) with temperature where the magnetic phases differ by an alignment of spins structure in Fe-Os layers.\cite{Paul,Kanungo1,Avijit} In present Sr$_2$FeIrO$_6$, the transition around 120 K is very weak, however, it needs a detailed investigation using microscopic experimental tools (neutron diffraction measurements discussed later). While previous studies report a single AFM transition (120 K) in Sr$_2$FeIrO$_6$, we show for the first time that the possibility of double magnetic transition in this material.\cite{Kharkwal} It remains, however, interesting that with increasing structural distortion or $x$, the transition at low temperature is no more evident while the high temperature transition becomes prominent.

\begin{figure}
\centering
		\includegraphics[width=8cm]{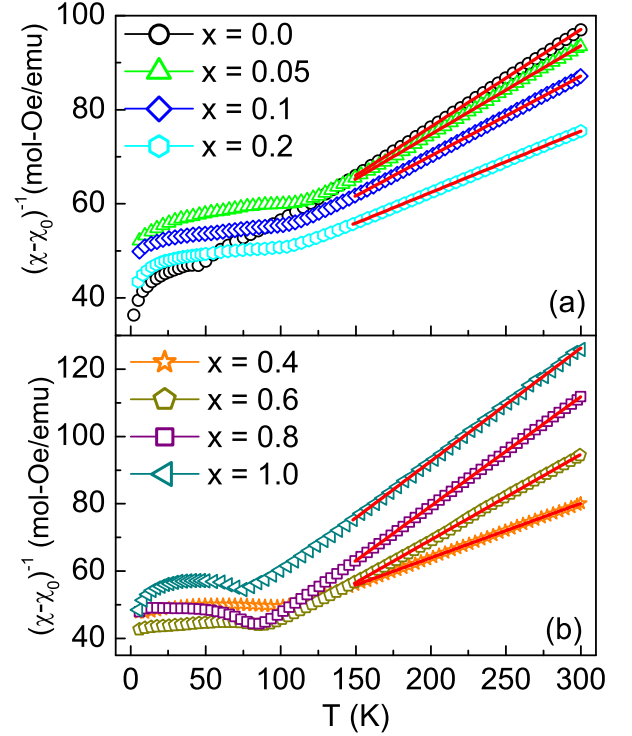}
\caption{(color online) A modified inverse magnetic susceptibility ($\chi$ - $\chi_0$)$^{-1})$ deduced from ZFC magnetization data are shown as a function of temperature in (a) $x$ = 0.0, 0.05, 0.1, 0.2 and (b) $x$ = 0.4, 0.6, 0.8, 1.0 for (Sr$_{1-x}$Ca$_x$)$_2$FeIrO$_6$ series. Solid lines are due to the linear fit of the data using modified Curie-Weiss model (Eq. 2).}
	\label{fig:Fig6}
\end{figure}

The magnetic susceptibility $\chi$ (= $M/H$) data in high temperature PM regime have been analyzed with modified Curie-Weiss (CW) behavior, Inverse magnetic susceptibility ($\chi^{-1}$) data as a function of temperature for x = 0.0, 0.05, 0.10, 0.20 and 0.4, 0.6, 0.8, 1.0 of (Sr$_{1-x}$Ca$_x$)$_2$FeIrO$_6$ series are shown in figures 6a and 6b respectively , where in the paramagnetic state $\chi^{-1}$ for all sample follows modified Curie-Weiss (CW) behavior. In the high temperature range data has been fitted well using the equation:

\begin{equation}
\chi = \chi_{0} +\frac{C}{T-\theta_{P}}
\end{equation}

where $\chi_0$ is the temperature independent contribution to the susceptibility which originates from core diamagnetism or paramagnetic moment, $C$ is the Curie constant and $\theta_P$ is the CW temperature. The effective PM  moment $\mu_{eff}$ has been calculated from $C$. The corrected inverse susceptibility ($\chi$ - $\chi_0$)$^{−1}$ as a function of temperature are shown in Figs. 6a and 6b for present series where the straight line fitting in PM regime (150 - 300
K) demonstrates the CW behavior (Eq. 2) in PM state. The parameters $\theta_P$ and $\mu_{eff}$ are shown in Figs. 8b and 8c, respectively as a function of doping concentration $x$ (discussed later).

\begin{figure}
\centering
		\includegraphics[width=8cm]{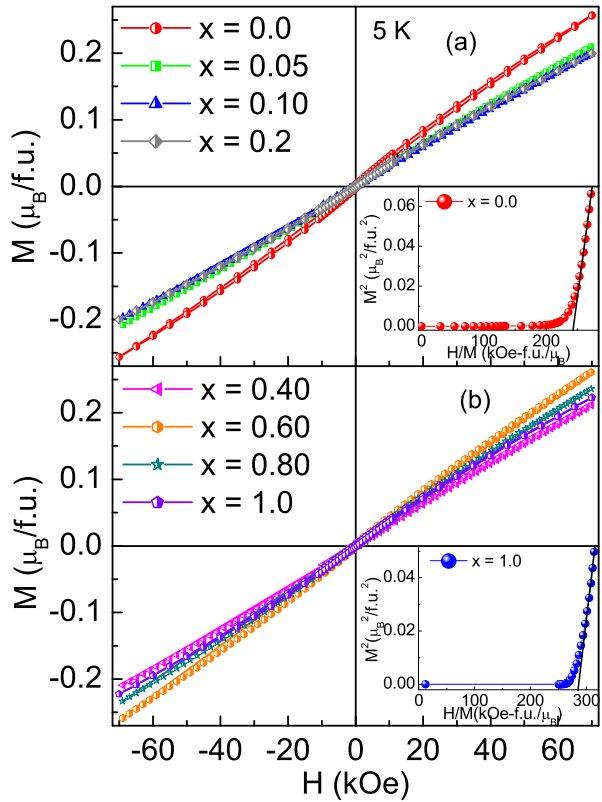}
\caption{(color online) Magnetic field dependent magnetization data $M(H)$ collected at 5 K are shown for (a) $x$ = 0.0, 0.05, 0.1, 0.2 and (b) $x$ = 0.4, 0.6, 0.8, 1.0 of (Sr$_{1-x}$Ca$_x$)$_2$FeIrO$_6$ series. Insets of (a) and (b) show the Arrott (M$^{2}$ vs $H/M$) plot for representative $x$ = 0.0 and 1.0, respectively.}
	\label{fig:Fig7}
\end{figure}

The magnetic field dependent isothermal magnetization $M(H)$ data collected at 5 K are shown in Figs. 7a and 7b for $x$ = 0.0, 0.05, 0.1, 0.2 and 0.4, 0.6, 0.8, 1.0, respectively of (Sr$_{1-x}$Ca$_x$)$_2$FeIrO$_6$ series. For $x$ = 0.0, the $M(H)$ data are almost linear till about 45 kOe, and above this field the $M(H)$ show a slight deviation from linearity with field. It is, however, evident in figure that $M(H)$ data do not exhibit saturation for all the samples up to maximum applied field i.e., 70 kOe. This linearity in the $M(H)$ data indicates a dominant AFM-type interaction in these materials. We calculate the magnetic moment at 70 kOe ($\mu_H$) to be 0.26 $\mu_B$/f.u. for $x$ = 0.0 which appears sufficiently small compared to its calculated value 5 $\mu_B$/f.u. (= $gS \mu_B$, g is Lande g-factor) taking contribution of Fe$^{3+}$ ($S$ = 5/2) and Ir$^{5+}$ ($S$ = 0). The experimentally determined $\mu_H$ has been shown in Fig. 8e with variation of $x$. The variation $\mu_H$ with $x$ where it shows an initial dip at $x$ $\sim$ 0.2 is quite consistent with behavior of $\theta_P$ in Fig. 8b. An increased value of $\left| \theta_P \right|$ promotes stronger AFM exchange across $x$ $\sim$ 0.2 which results in a reduced moment $\mu_H$ even in presence of magnetic field.

Further, a small hysteresis in the $M(H)$ data has been observed with coercive field H$_c$ $\sim$ 590 Oe and an remnant magnetization M$_r$ $\sim$ 2.8 $\times$ 10$^{-3}$ $\mu_B$/f.u for parent Sr$_2$FeIrO$_6$ ($x$ = 0.0) material. The respective H$_c$ values for present series are shown in Fig. 8f. The H$_c$ values, after showing a initial dip maximizes for $x$ in between 0.6 to 0.8. The remnant magnetization M$_r$ also follows similar trend where it first decreases from 2.8 $\times$ 10$^{-3}$ to 2.1 $\times$ 10$^{-3}$ $\mu_B$/f.u for $x$ = 0.2 and then increases to 3.75 $\times$ 10$^{-3}$ $\mu_B$/f.u for $x$ = 0.6 and again decreases to 2.16 $\times$ 10$^{-3}$ $\mu_B$/f.u. This H$_c$ and M$_r$ rather follow the similar tend of $\theta_P$ with $x$. The nature of magnetism in these materials is further checked with Arrott’s plot ($M^2$ vs $H/M$ ),\cite{Arrott} as shown in inset of Figs. 7a and 7b for representative $x$ = 0.0 and 1.0, respectively. A positive intercept due to linear fitting in high field regime of Arrott plot signifies a FM nature of the system whereas the negative intercept excludes the possibility of FM interactions. Both insets of Fig. 7 indicate a negative intercept in the Arrott plot which implies a non-FM type exchange interaction in the present series in confirmation with the other results.

\begin{figure}
\centering
		\includegraphics[width=6cm]{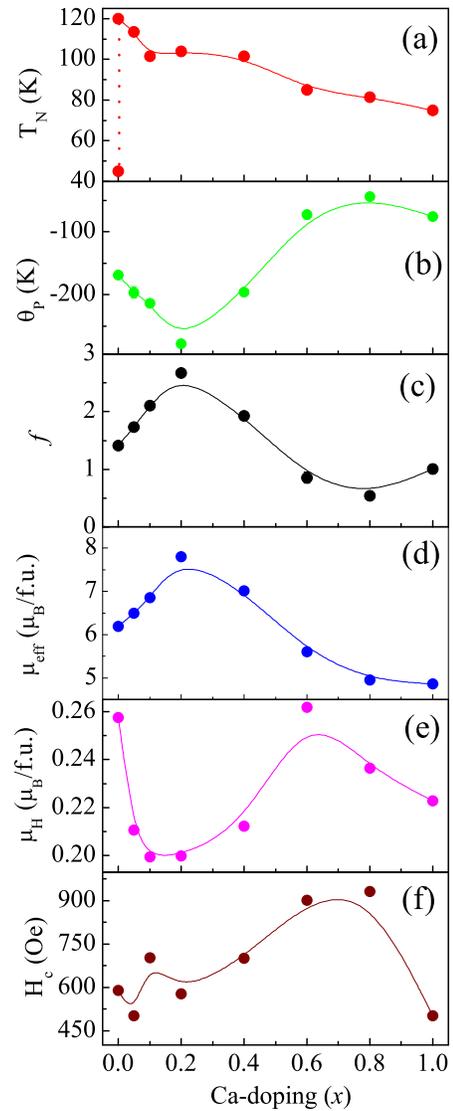}
\caption{(color online) (a) $T_N$, (b) $\theta_P$, (c) frustration parameter $f$ (d) $\mu_{eff}$, (e) $\mu_H$, (f) $H_c$ are shown as a function of doping concentration $x$ for (Sr$_{1-x}$Ca$_x$)$_2$FeIrO$_6$ series. Lines are guide to eyes.}
	\label{fig:Fig8}
\end{figure}

The magnetic parameters of present series are noted in Fig. 8. The AFM transition temperature $T_N$ shows almost a continuous decrease across the series (Fig. 8a), however, the low temperature AFM transition (T$_N$ $\sim$ 45 K) that is observed in $x$ = 0 material, is not evident in Ca doped ones. However, the interesting effect of isoelectronic Ca substitution in present series is observed in other magnetic parameters. For Sr$_2$FeIrO$_6$, the $\theta_P$ is found to be -169 K. While the sign $\theta_P$ is an indicative of non-FM or AFM type magnetic exchange interaction, its magnitude suggests strength of interaction is reasonably strong. As evident in Fig. 8b, while keeping the sign same the magnitude of $\theta_P$ initially increases till around $x$ = 0.2 reaching its value -278 K, and then it decreases. This initial increase of $\theta_P$ is quite interesting as T$_N$ shows a continuous decrease. This implies the strength of AFM exchange interaction increases with an initial substitution of Ca till $x$ $\sim$ 0.2. Nonetheless, the higher magnitude of $\theta_P$ than the transition temperature T$_N$ implies a magnetic frustration in the system. The calculated frustration parameter ($f$ = $\theta_P$/T$_N$), which tentatively measures the level of frustration, has been shown in Fig. 8c for present series. The $f$ $>$ 1 for parent Sr$_2$FeIrO$_6$ indicates a magnetic frustration is present in this material, which further increases till $x$ $\sim$ 0.2 indicating frustration increases with initial substitution of Ca. The $f$, however, decreases for $x$ $>$ 0.2 while its value $\sim$ 1 for Ca$_2$FeIrO$_6$ suggests an agreement between ordering temperature and strength of magnetic interaction. The double perovskites, in deed, have inherent geometrical frustration arising from its structural organization. As seen in unit cell of Sr$_2$FeIrO$_6$ (Fig. 9b), Fe and Ir atoms occupy alternate position. The Fe/Ir sublattices form different set of tetrahedra where the vertices are occupied by Fe/Ir atoms. If the occupying elements are magnetic in nature and they are engaged in AFM type magnetic interactions, it generates magnetic frustration in the system. Here, it can be mentioned that although present systems exhibit $f$ $>$ 1 but its values are much lower than the value usually seen for frustrated systems ($f$ $>$ 10).

\begin{figure*}
\centering
		\includegraphics[width=16cm]{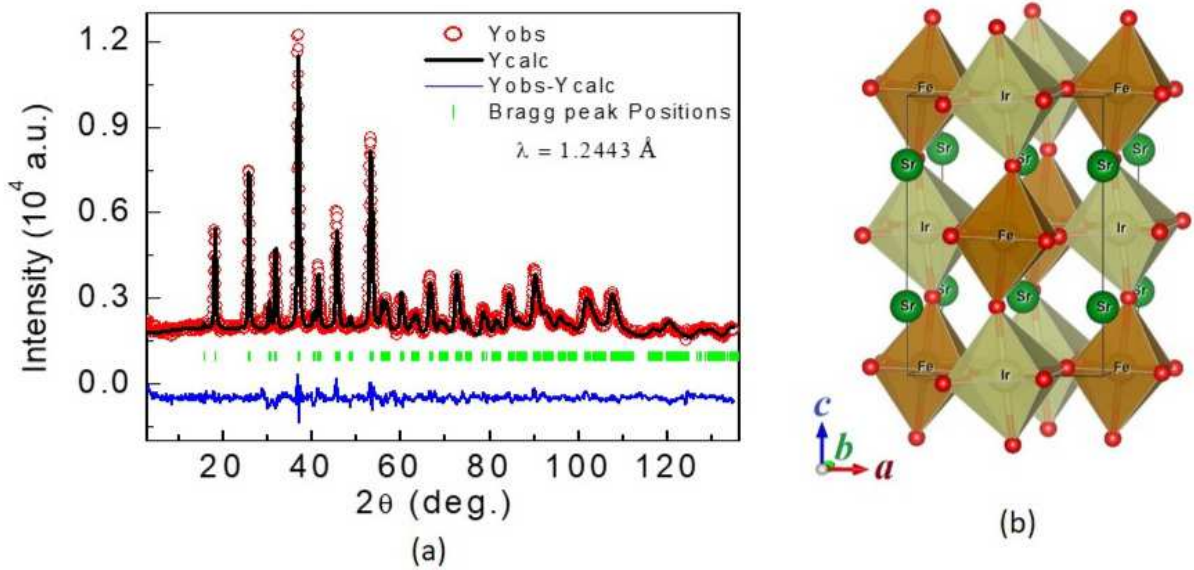}
\caption{(color online) The experimentally observed (circles) and calculated (solid lines through the data points) neutron diffraction patterns for Sr$_2$FeIrO$_6$ at room temperature measured by PD-2 ($\lambda$ = 1.2443 \AA). The solid lines at the bottom of the each panel represent the difference between observed and calculated patterns. The vertical bars indicate the positions of allowed nuclear Bragg peaks. (b) The crystal structure of Sr$_2$FeIrO$_6$}
	\label{fig:Fig9}
\end{figure*}

The effective magnetic moment $\mu_{eff}$, on the other hand, shows an opposite behavior of $\theta_P$, showing its maximum value for $x$ = 0.2 material (Fig. 8d). The $\mu_{eff}$ has been calculated theoretically using the formula $\left(\mu_{eff} =\sqrt{(\mu_{Fe})^{2} +(\mu_{Ir})^{2}}\right)$, where $\mu_{Fe}$ and $\mu_{Ir}$ are the magnetic moment of Fe$^{+3}$ and Ir$^{+5}$ cations, respectively. For Sr$_2$FeIrO$_6$, $\mu_{eff}$ is found to be close to theoretically calculated value with Fe$^{3+}$ and Ir$^{5+}$ charge states, where Fe$^{3+}$ ($S$  =  5/2) only contributes to the moment while Ir$^{5+}$ ($J_{eff}$ = 0) is considered to be nonmagnetic. Under the picture of cubic crystal field and strong SOC, spins of Ir$^{5+}$ (5$d^4$) completely occupy the $J_{eff}$ = 3/2 state with zero effective magnetic moment. On contrary, if one considers Fe$^{3+}$ is in $S$ = 5/2 high-spin state and Ir$^{5+}$ is in $S$ = 1 low-spin state under the strong distortion driven splitting, then in addition to spin moment there would be orbital contribution to the moment at Ir-site. Considering the both orbital and spin magnetic moment of Fe and Ir, the total calculated moment comes out around 6.41 $\mu_B$/f.u, which is quite close to the experimentally determined value of $\mu_{eff}$ (Fig. 8d). Although the bulk magnetization measurements are not quite sensitive to the orbital moment part, if we consider only spin moment of Fe and Ir the ideal magnetic moment comes out to be $\sim$ 6.56 $\mu_B$, that is also close to the measured $\mu_{eff}$ values. Here point to be noted that, spectroscopic studies do not find any evidences of charge state changes with Ca at Fe/Ir sites (Fig. 4). Moreover, 3$d$-Fe$^{3+}$ and 5$d$-Ir$^{5+}$ usually prefer to stay in high- and low-spin state, respectively considering the six coordinated octahedral environment of inorganic ligand and resulting competition between the crystal field and the electron-electron Coulomb correlation. Therefore, we tried to understand the spin state through the electronic structure calculations presented afterwards. It is clearly understood that the substitution of isoelectronic Ca at Sr-site will unlikely change the Fe/Ir charge state but would introduce a local distortion in respective Fe/Ir octahedra. This would influence the crystal field effect, hence the overall magnetic behavior. For instance, Cao \textit{et al.}\cite{Cao} have discussed the possibility of different magnetic states of Ir$^{5+}$ ($S$ = 0 and 1) in DP Sr$_2$YIrO$_6$ which mainly originates due to an interplay between SOC, (noncubic) crystal fields and local magnetic exchange interactions. However, this noncubic crystal field vary from material to material based on their lattice distortion.

\begin{table}
\caption{\label{tab:table 1} The Rietveld refined lattice constant ($a$, $b$, $c$), fractional atomic coordinates, and isotropic thermal parameters ($B_{iso}$) for Sr$_2$FeIrO$_6$ at 300 K.}
\begin{ruledtabular}
\begin{tabular}{cccc}
300 K\\
\hline
Space group = \textit{I$\bar{1}$},& $a$ = 5.558(1)\AA,& $b$ = 5.569(1)\AA,& $c$ = 7.8082(6)\AA, \\
  $\alpha$ = 90.30(1)$^{\circ}$,& $\beta$ = 89.37(1) $^{\circ}$,& $\gamma$ = 89.92(1)$^{\circ}$,& $V$=241.72(6) \AA$^{3}$\\
	\hline
Atom & x/a & y/b & z/c \\
\hline
Sr & 0.5009  & 0.4975 & 0.2520\\
Fe &  0.0 & 0.5 & 0.0\\
Ir & 0.5 & 0.0 & 0.0\\
O1 & 0.2520 & 0.2612 & -0.0038\\
O2 & 0.2245 & 0.7799 & -0.0162\\
O3 & 0.4554 & 0.0094 & 0.2479\\
\hline
\end{tabular}
\end{ruledtabular}
\end{table}
 
\subsection{Neutron powder diffraction study on Sr$_2$FeIrO$_6$}
To understand the crystal structure down to low temperature as well as to find out the magnetic structure in Sr$_2$FeIrO$_6$ ($x$ = 0), neutron powder diffraction (NPD) measurements have been carried out at 300, 150 and 5 K which represent both nonmagnetic and magnetic states. The crystal structure for Sr$_2$FeIrO$_6$ is found to be triclinic with space group \textit{I$\bar{1}$} over the studied temperature range 5-300 K and is consistent with the structure obtained from the XRD data (Fig. 2). In our Rietveld analysis, we obtain lower $\chi^2$ value with triclinic-$I\bar{1}$ symmetry (4.92) compared to monoclinic-$I2/m$ structure (5.26), however, the difference is not very significant. Nonetheless, the analysis indicate a distorted crystal structure (triclinic/monoclinic) which is also consistent with multiple peaks in diffraction pattern (Fig. 9a). The Rietveld refinement of neutron diffraction pattern at room temperature is shown in Fig. 9a. The refined structural parameters are given in Table 1. The schematic representation of the crystal structure of Sr$_2$FeIrO$_6$ is shown in Fig. 9b. Our Rietveld analysis also indicates a relatively small anti-site disorder of about 9(2)\% among the Ir and Fe sites in the parent Sr$_2$FeIrO$_6$ compound. We hope that similar trend of anti-site disorder will be retained across the series. Given that the degree of anti-site disorder depends on the difference between ionic radii and the charge state of involved transition metal cations and this has large impact on the physical properties of materials, the estimated disorder in present systems is comparable to 3$d$-5$d$ based other double perovskite systems.\cite{Narayanan,Kayser1}  

\begin{figure*}
\centering
		\includegraphics[width=16cm]{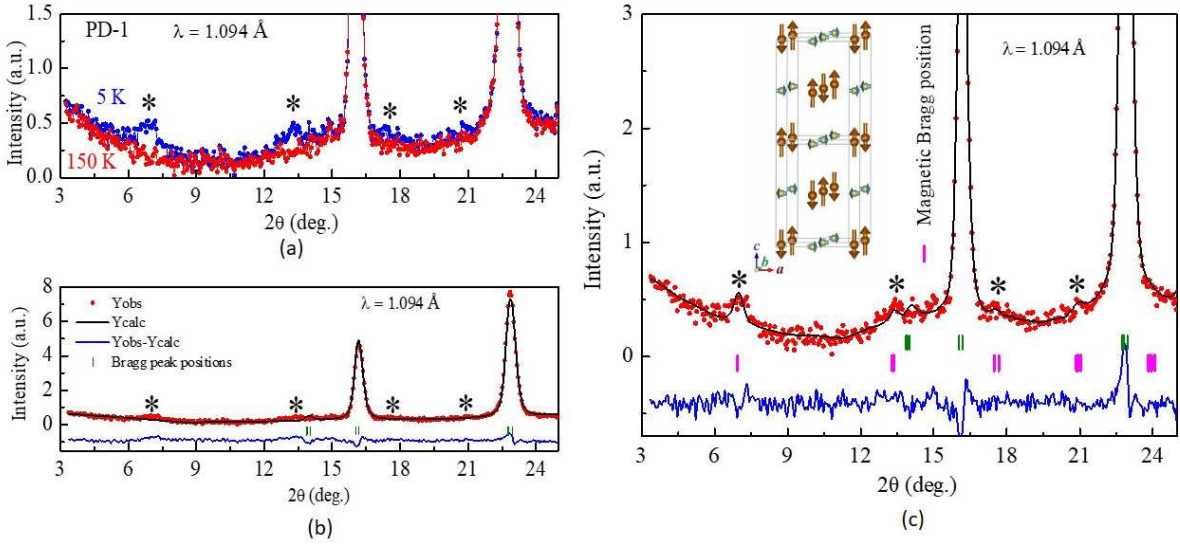}
\caption{(color online) The neutron diffraction patterns for Sr$_2$FeIrO$_6$ measured at 5 and 150 K by PD-1 ($\lambda$ = 1.094 \AA). The patterns are zoomed over the lower angular range to show the magnetic Bragg peaks. The magnetic peaks at 5 K are marked by asterisks. (b) The Rietveld refined (considering only the nuclear phase) neutron diffraction pattern measured at 5 K. (c) The Rietveld refined (considering nuclear and magnetic phases) neutron diffraction pattern measured at 5 K. The vertical bars indicate the positions of allowed nuclear (top panel) and magnetic (bottom panel) Bragg peaks. The inset shows the derived magnetic structure for Sr$_2$FeIrO$_6$.}
	\label{fig:Fig10}
\end{figure*}

The neutron diffraction patterns measured at 5 (magnetically ordered state) and 150 K (paramagnetic state) are shown in Fig. 10a. Appearance of additional weak magnetic Bragg peaks at $\sim$ 6.9$^{\circ}$ and 13.4$^{\circ}$  are evident in the diffraction pattern measured at 5 K, confirming an antiferromagnetic ordering in Sr$_2$FeIrO$_6$. These two magnetic Bragg peaks could be indexed with a propagation vector $k$ = (0, 0.5, 0.5) with respect to the triclinic nuclear unit cell. The symmetry-allowed magnetic structure is determined by a representation analysis using the program BASIREPS available within the FULLPROF suite.\cite{Full} The symmetry analysis reveals that there is only one possible magnetic structure. The two phase (nuclear + magnetic) refinement of the measured diffraction pattern at 5 K is shown in Fig. 9c. A good agreement between observed and calculated pattern is evident ($R_{mag}$ $\sim$ 17\%). The corresponding magnetic structure is shown in the inset of Fig. 9c which is found to be pure antiferromagnetic in nature without having any net magnetization per unit cell. At 5 K, the site ordered moments of Fe and Ir ions are found to be $\sim$ 4.5(2) and 0.5(3) $\mu_B$/site, respectively. The Fe and Ir moments are aligned along the $c$ and $a$ axes, respectively, i.e., the moments are orthogonal to each other. The derive magnetic ground state is in agreement with the same obtained by DFT calculation (discussed later). The obtained magnetic structure is further in good agreement with that, as reported by Kayser \textit{et al.}\cite{Kayser}

\subsection{Electronic transport in (Sr$_{1-x}$Ca$_x$)$_2$FeIrO$_6$}
The electrical resistivity ($\rho$) as a function of temperature are shown in Fig. 11 for (Sr$_{1-x}$Ca$_x$)$_2$FeIrO$_6$ series. From the figure, it is evident that the resistivity increases sharply with decrease in temperature. At very low temperature the resistivity increases by several orders of magnitude than its room temperature value. Moreover, an insulating nature of charge transport is apparent throughout the temperature range for whole series. For $x$ = 0.0, we observe a change in slope of $\rho (T)$ data across its low temperature magnetic transition around 45 K.\cite{Kharkwal} The evolution of resistivity with composition in present series is though nonmonotonic. The left inset of Fig 11 shows the resistivity value at 40 K for varying Ca composition. As seen in figure, the resistivity initially shows a peak at $x$ $\sim$ 0.1 and then its value increases by several orders and shows another peak at $x$ $\sim$ 0.6. This nonmonotonic variation of $\rho(x)$ mimics that of structural parameters which also show an anomalous change across $x$ = 0.1 and 0.6 (Fig. 3). Therefore, Figs. 3, 8 and 11 together suggest that the evolution of structural, magnetic and electrical transport behavior in present series are correlated.

This series of samples show a highly  insulating  state where the effect of magnetic field on charge transport has been checked at low temperature. The right inset of Fig. 11 shows the magnetoresistance (MR), expressed as $\left[\Delta \rho/\rho_{0} (\%)= (\rho-\rho_{0}) /\rho_{0} \times 100\right]$ for $x$ = 1.0 at 25 K. A negative MR has been observed till the maximum applied field of 30 kOe. The observed MR is almost linear and shows a little change (-3\%) with field. Usually, the materials with sizable SOC effect display positive MR induced by weak antilocalization (WAL) effect. The prominent examples are Bi$_2$Se$_3$,\cite{chen1} Bi$_2$Ti$_3$,\cite{he} and even in iridate Na$_2$IrO$_3$ films.\cite{jenderka} However, the negative MR at low temperatures is mainly understood in terms of weak localization (WL) effect, hopping conduction and related magnetic scattering effect. This is related to the quantum interference (QI) effect which is largely viewed as the quantum correction in the classical Drude model for electronic conduction. In case of hopping mediated conduction, the probability of hopping between two occurring sites depends on an interference of connecting paths. The effect of magnetic field leads to a destruction of QI effect producing a negative MR. Following an approach of averaging the logarithm of conductivity over many random paths, it has been shown that the effect of magnetic field is linear over charge conduction, showing a linear negative MR.\cite{nguyen}

\begin{figure}
\centering
		\includegraphics[width=8cm]{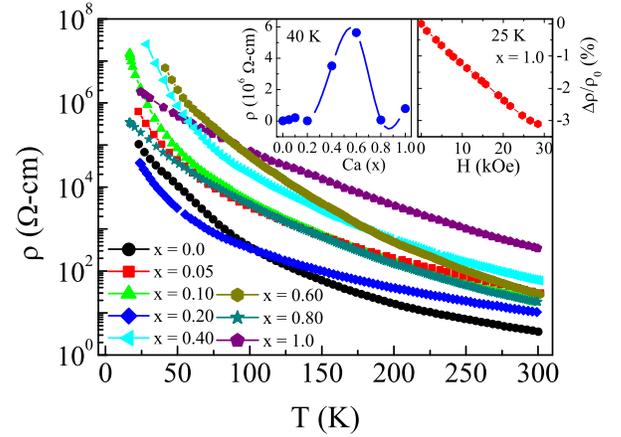}
\caption{(color online) Resistivity data as a function of temperature are shown in semi-log scale for (Sr$_{1-x}$Ca$_x$)$_2$FeIrO$_6$ series. Left inset shows the resistivity value at 40 K against Ca doping showing a peak around $x$ = 0.1 and 0.6. The line is a guide to eyes. Right inset shows the MR data collected at 25 K for $x$ = 1.0 material till 30 kOe.}
	\label{fig:Fig11}
\end{figure}

In our previous study, we have shown that the electron conduction in Sr$_2$FeIrO$_6$ DP follows Mott's 3-dimensional (3D) variable range hopping (VRH) model. With the substitution of Ca at Sr-site, we do not find any change in the electron conduction behavior. The electron conduction mechanism over the series can be described well with 3D VRH model,\cite{Mott}

\begin{equation} 
\rho = \rho_{0} \exp{\left[\left(\frac{T_0}{T}\right)^{4}\right]}
\end{equation}

where T$_0$ is the characteristics temperature which is given as following,

\begin{equation} 
T_0 = \frac{18}{k_{B} N(E_{F})\xi^{3}}
\end{equation}
 
where $k_B$ is the Boltzmann constant and N(E$_F$), $\xi$ are density of states (DOS) at the Fermi surface and localization length, respectively. Figs. 12a and 12b show $\log(\rho)$ vs $T^{-1/4}$ plot of the $\rho(T)$ data following Eq. 3 for $x$ = 0.0, 0.05, 0.1, 0.2 and 0.4, 0.6, 0.8, 1.0 of (Sr$_{1-x}$Ca$_x$)$_2$FeIrO$_6$ series, respectively. We have previously shown that the $\rho(T)$ of Sr$_2$FeIrO$_6$ can be fitted in two different temperature ranges with a crossover around 45 K ($T_N$). Given that N(E$_F$) does not change appreciably over temperature in insulators, this change in slope ($T_0$) across the magnetic transition temperature can be through change in localization length (Eq. 4). As evident in Fig. 12, the $\rho(T)$ can be fitted with Eq. 3 in two different linear regimes till $x$ = 0.6, however, $x$ = 0.8 and 1.0 materials the Eq. 3 is obeyed only in high temperature regime. Different regions in the fitting of $\rho(T)$ data closely follow the magnetic transitions in these materials. The temperature range and the characteristics temperature T$_0$ obtained from the fitting of the resistivity data are shown in Table II. 

\begin{figure}
\centering
		\includegraphics[width=9cm]{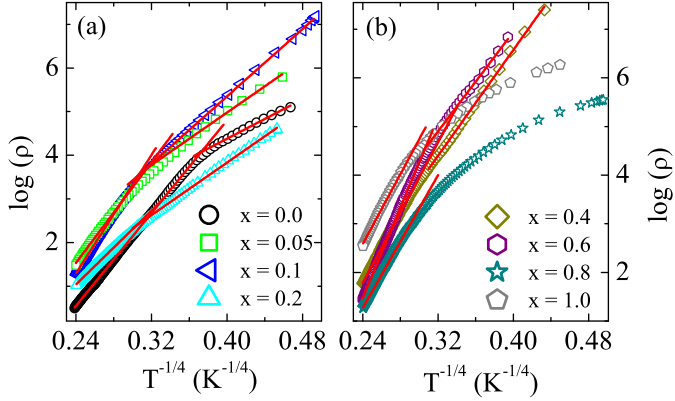}
\caption{(color online) $\log(\rho)$ vs T$^{-1/4}$ plot of resistivity data following Eq.3 of VRH model are shown for (a) x = 0.0, 0.05, 0.1, 0.2 and (b) x = 0.4, 0.6, 0.8, 1.0 of (Sr$_{1-x}$Ca$_x$)$_2$FeIrO$_6$ series. Red solid line is due to fit of the resistivity data.}
	\label{fig:Fig12}
\end{figure}

\begin{table}
\caption{\label{tab:table 2} Temperature range and characteristics temperature (T$_0$) obtained using Eq. 4 from the fitting of resistivity data are shown with doping concentration ($x$) for (Sr$_{1-x}$Ca$_x$)$_2$FeIrO$_6$ series.}
\begin{ruledtabular}
\begin{tabular}{ccc}
Sample ($x$) & Temperature range (K) & T$_0$ (10$^5$ K)\\ 
\hline
0.0 & 300 - 52 & 0.16\\
\space & 45 - 20 & 4.93\\
0.05 & 300 - 125 & 6.9\\
\space & 60 - 22 & 0.51 \\
0.10 & 300 - 95 & 13.3\\
\space & 83 - 20 & 1.38 \\
0.2 & 300 - 95 & 2.0 \\
\space & 84 - 24 & 0.51 \\
0.4 & 300 - 122 & 17.02 \\ 
\space & 82 - 28 & 5.46 \\
0.6 & 300 - 122 & 46.43\\
\space & 72 - 41 & 4.27 \\
0.8 & 300 - 140 & 13.6 \\
\space & 122 - 65 & 1.51 \\
\space & 35 - 16 & 0.18 \\
1.0 & 300 - 110 & 16.57 \\
\space & 105 - 66 & 9.4 \\
\space & 65 - 24 & 0.63
\end{tabular}
\end{ruledtabular}
\end{table}

As evident in Table II, the change of T$_0$ parameter with composition is nonmonotonic. Further, the change in T$_0$ with temperature for a specific material is likely due to increase or decrease in the localization length $\xi$ since the N(E$_F$) would not change significantly in case of insulators. Nonetheless, as the resistivity, crystal structure and magnetism exhibit large change with doping concentration $x$, therefore we believe that both N(E$_F$) and $\xi$ contribute for the modification of T$_0$.

\subsection{Electronic structure calculations}
\begin{figure*}
	\centering
		\includegraphics[width=16cm]{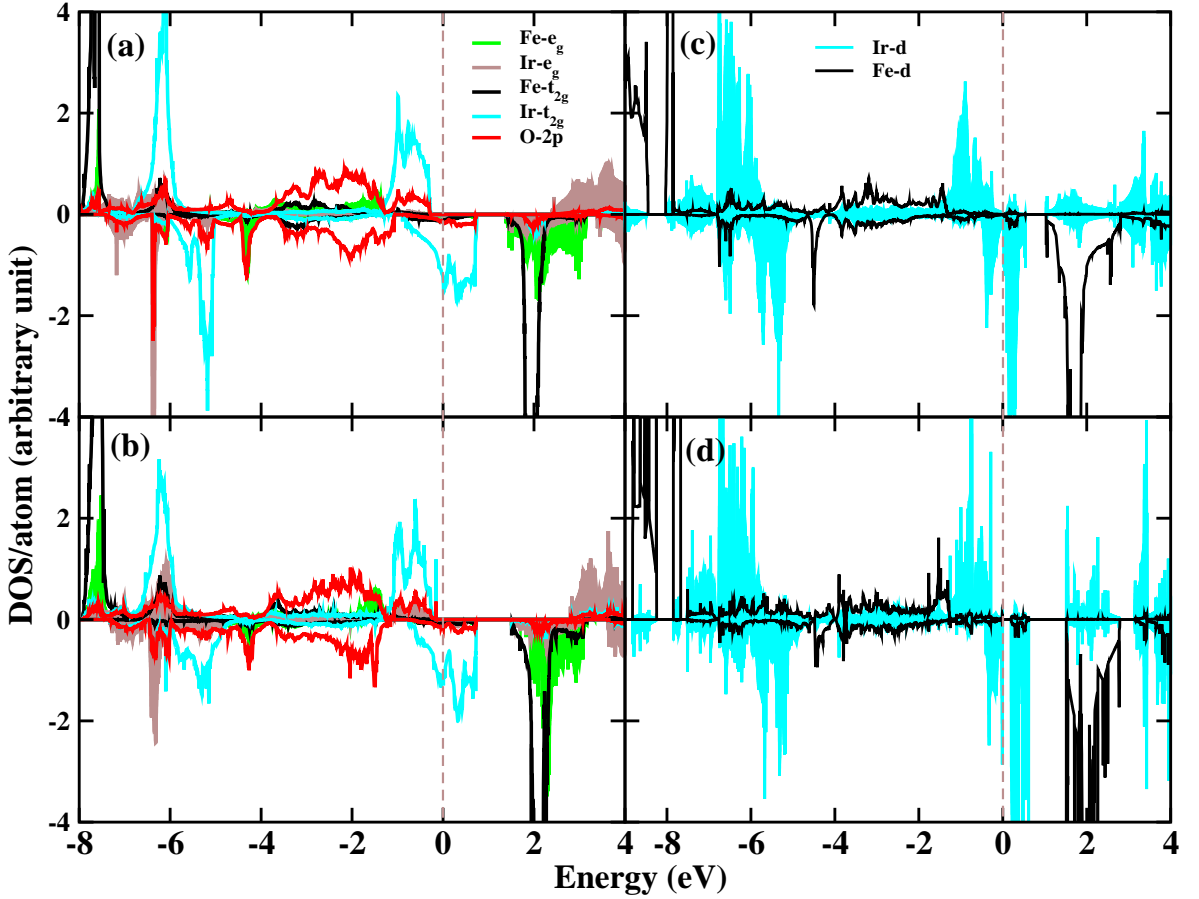}
	\caption{The calculated GGA+$U$ orbital projected DOS for the FM state are shown for (a) Sr$_2$FeIrO$_6$ and (b) Ca$_2$FeIrO$_6$. The (c) and (d) show the similar GGA+$U$ DOS calculated for the energetically lowest AFM state for Sr$_2$FeIrO$_6$ and Ca$_2$FeIrO$_6$, respectively. The used $U$ values are, $U^{Fe}_{eff}$ = 5 eV and $U^{Ir}_{eff}$ = 2 eV. The Fermi energy is set at zero in the energy scale.}
	\label{fig:Fig13}
\end{figure*}

To understand the observed experimental results in the present series of materials from microscopic point of view, we have done the electronic structure calculations at the DFT level, by taking care of electronic correlation (GGG+$U$) through Hubbard $U$ needed for transition metals as well as by considering spin-orbit coupling (GGG+$U$+SOC) required for the 5$d$ elements. We have done calculations with different spin configurations. The calculated electronic density of states (DOS) through GGA+$U$ formalism for both ferromagnetic (FM) and antiferromagnetic (AFM) spin alignments are shown in Fig. 13. For all the calculations, we have used $U^{Fe}_{eff}$ = 5 eV and $U^{Ir}_{eff}$ = 2 eV, where $U_{eff} = U - J_H$ ($J_H$ is the Hund's coupling, set to be 0.9 eV) which suits well according to 3$d$ and 5$d$ elements. The GGA+$U$ calculated DOS for the FM spin alignments turn out to be half-metallic in nature with a finite DOS at the Fermi level in the minority spin channel coming from the Ir-$t_{2g}$ states, while the majority spin channel shows a gap for the both Sr$_2$FeIrO$_6$ (SFIO) and Ca$_2$FeIrO$_6$ (CFIO), as shown in the Figs. \ref{fig:Fig13}a and \ref{fig:Fig13}b, respectively.
 
Interestingly, the calculated GGA+$U$ DOS with exactly the same values of Hubbard $U$ (mentioned earlier), but with the AFM spin configuration, instead of FM configuration, is able to open a gap in both the spin channels by splitting the minority Ir-$t_{2g}$ bands for both SFIO and CFIO, as shown in Figs. \ref{fig:Fig13}c and \ref{fig:Fig13}d, respectively. Although the gaps at the Fermi level are very tiny (0.02 eV for SFIO and 0.2 eV for CFIO), however, only decent Hubbard $U$ along with AFM spin correlation is able to open a gap in electronic band structure over the entire Brillouin zone, resulting an insulating state for the both SFIO and CFIO. Given that present (Sr$_{1-x}$Ca$_x$)$_2$FeIrO$_6$ series exhibits an insulating and AFM ground state (see Figs. \ref{fig:Fig5}, \ref{fig:Fig7} and \ref{fig:Fig11}), our calculations indeed agree with the experimental evidences and indicate that the investigated materials falls under the category of weak Mott-type AFM insulator, as the electronic correlations for the 5$d$-Ir is comparatively weaker. Also, we understand that the enhanced gap at the Fermi level in CFIO compared to the SFIO in the GGA+$U$ calculations with an AFM state (Figs. \ref{fig:Fig13}c and \ref{fig:Fig13}d), is due to an enhanced structural distortion in CFIO compared to that in SFIO. Point to be further noted that, a gapped insulating ground state can also be realized even with FM correlation for an unphysical high value of Hubbard $U$ at the 5$d$-Ir elements (i.e., $U^{Ir}_{eff}$ $\geqslant$ 4 eV), which is due to the splitting of Ir-$t_{2g}$ manifold into lower and upper Hubbard like bands, very similar to that of the typical correlation driven Mott-insulator cases. 

The closure investigation of the orbital projected DOS, as shown in Figs. \ref{fig:Fig13}a and \ref{fig:Fig13}b, reveal that due to the distorted octahedral environment, the Fe-$d$ states are broadly split into $t_{2g}$-$e_g$ states, where both of the states, are completely filled in the majority spin channel and completely empty in the minority spin channel, for both SFIO and CFIO. Similarly, the Ir-$d$ states are split into low lying $t_{2g}$ and high lying $e_g$ states, however, the Ir-$t_{2g}$ states are completely filled in majority spin channel and partially filled in minority spin channel having dominant contribution near to the Fermi level. The Ir-$e_g$ states, on the other hand, are completely empty in both the spin channels lying 3 to 4 eV above the Fermi level in the conduction band. The calculated magnetic moment (Fe = 4.28 $\mu_B$/site and Ir = 1.07 $\mu_B$/site) and DOS suggest that iron is in Fe$^{3+}$ (3$d^5$) nominal valence state whereas iridium is in Ir$^{5+}$ (5$d^4$) nominal valance state with high spin ($S$ = 5/2) and low spin ($S$ = 1) state for Fe and Ir, respectively for both SFIO and CFIO. The calculated magnetic moments and the valance states support the experimental results, as seen in the XPS and Neutron measurements. The substantial hybridization between Fe/Ir-$d$ and O-$2p$ states are evident form the DOS as well as from induced magnetic moment at oxygen sites (average moment 0.06 $\mu_B$/site). The reduced moment at the Ir sites compared to that of the ideal value are due to the substantial orbital contribution and hybridization effect which will be discussed in the next section.

\begin{figure}
	\centering
		\includegraphics[width=7.9cm]{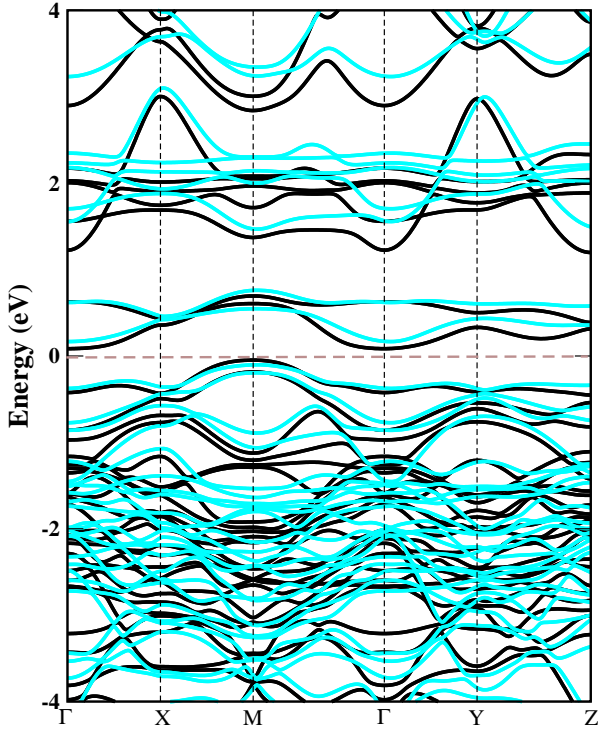}
	\caption{The calculated GGA+$U$+SOC electronic band structure of AFM state are shown for SFIO (black) and CFIO (cyan). The Fermi energy is set at zero in the energy scale.}
	\label{fig:Fig14}
\end{figure}

To understand the influence of spin-orbit coupling (SOC), the electronic band structure of SFIO and CFIO have been calculated taking combination of GGA+$U$+AFM+SOC along the high symmetry direction of Brillouin zone (BZ) after fixing the spin quantization axis along [001] (Fig. \ref{fig:Fig14}). Due to the introduction of SOC, the degeneracy of the bands is broken, however, there are no such drastic difference between SFIO and CFIO band structure. As evident from the band structure in Fig. \ref{fig:Fig14}, an indirect band gap between $\Gamma$ and $M$ point in the BZ, which is slightly higher in CFIO than that in SFIO due to increased structural distortion caused by the smaller cationic size of Ca. Although, the gap for both of the compounds has enhanced due to the inclusion of SOC, compared to the the GGA+$U$ calculations, however, the enhancement is more prominent in case of SFIO than the CFIO, as the structural distortion is larger for the CFIO compared to that of the SFIO, which competes with the strength of the SOC. Also, there are no such drastic changes found in the band structure between GGA+$U$ and GGA+$U$+SOC calculations, except the lifting of band degeneracy at few selected K-points and the enhancement of the band gap. The calculations show that the Fe spin magnetic moment minimally changes after the inclusion of SOC for the both SFIO and CFIO, with a small orbital magnetic moment emerges which is expected due to the quenching of orbital degrees of freedom in $d^5$ high spin configuration. On the other side, the Ir spin magnetic moment [Ir-SFIO = 0.62 $\mu_B$/site, Ir-CFIO = 0.61 $\mu_B$/site] reduces substantially due to introduction of SOC. In addition to that, a large orbital magnetic moment around 0.25 $\mu_B$/site with same sign as that of the spin moment has been found for both SFIO and CFIO. Point to be noted here is that, although there are substantial orbital magnetic moments emerge at Ir site, still the orbital magnetic moment is smaller than the corresponding spin moment ($\mu^{orb}/\mu^{spin} \approx$ 0.5). This suggests that for present materials, the effective strength of the SOC is small compared to that of the conventional strong SOC-driven Mott insulators such as, Sr$_2$IrO$_4$, $A_2$Ir$_2$O$_7$, Na$_2$IrO$_3$,\cite{Kim,Kim1,Krempa,Rau,Yogesh,Pesin} where the ratio ($\mu^{orb}/\mu^{spin}$) is close to as high as 2. Beside that, our calculations indicate even though SOC is required to stabilize an AFM state and it enhances the band gap value, but SOC is not essentially required to open a gap in DOS or to induce an insulating state in present materials. This suggests these materials neither belong to the conventional correlation-driven Mott insulators nor the conventional SOC-driven Mott insulator ($J_{eff}$ states), rather can be classified as the `SOC enhanced correlation-driven AFM-Mott insulator'.

\begin{figure}
	\centering
		\includegraphics[width=8cm]{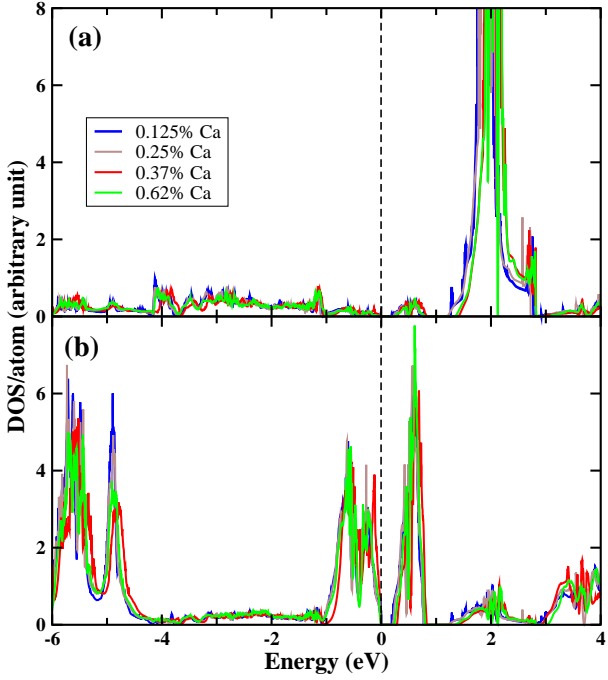}
	\caption{(a) Fe-$3d$ and (b) Ir-$5d$ electronic DOS, calculated under GGA+$U$+SOC scheme for AFM state, are shown for (Sr$_{1-x}$Ca$_x$)$_2$FeIrO$_6$ series with the composition ($x$) mentioned in the plot. The Fermi energy is set at zero in the energy scale.}
	\label{fig:Fig15}
\end{figure}

We have examined the stability of magnetic ground state (FM or AFM) from total energy perspective and have found that SOC is indeed needed to stabilize AFM ground state over the FM ordering for both SFIO and CFIO by an amount 2.35 meV/f.u and 25.56 meV/f.u, respectively. We have also investigated the energetics of different AFM spin configurations and the calculations show that in-plane ($ab$-plane) nearest neighbor Fe and Ir spins are coupled antiferromagnetically whereas out-of-plane ($c$-direction) nearest neighbor Fe and Ir spins are coupled ferromagnetically, which is consistent with the neutron diffraction measurements as shown in the inset of Fig.\ref{fig:Fig10}c. However, the interpenetrating two FCC lattices of Fe and Ir form a frustrated lattice where all the nearest neighbor interactions are not simultaneously satisfied, as evident from the both theoretical calculations as well as experimental results obtained from neutron diffraction. From the energy differences, it is clear that the magnetic exchange interactions between Fe and Ir sites are almost an order of magnitude stronger in the case of CFIO than that of the SFIO. In ordered DPs, the super-exchange interactions are mainly dominated via two pathways; one through nearest neighbor (NN) Fe-O-Ir and another through next nearest neighbors (NNN) i.e., Fe-O-Ir-O-Fe and/or Ir-O-Fe-O-Ir pathways, as governed by the Goodenough-Kanamori angular rule.\cite{goodenough,kanamori} However, as the size of $A$-site cation controls the super-exchange angle, the strength and sign of the exchange interactions are tuned accordingly. With Ca doping, our calculations indeed indicate that the super-exchange angles are changed substantially. For instance, in CFIO ($\sim$ 150$^o$) the angles are more deviated from the ideal 180$^o$, compared to SFIO ($\sim$ 166$^o$). For intermediate doping levels, the evolution of angles is nonmonotonic due to an uneven distortion in the local environment. The electronic structure total energy calculations clearly indicate that the AFM exchange interactions between different magnetic sites in CFIO is much stronger than that of the SFIO. These theoretical results are also in agreement with the experimentally determined ordering temperature T$_N$ of present series (Fig.\ref{fig:Fig7}).

\begin{figure}
	\centering
		\includegraphics[width=8.6cm]{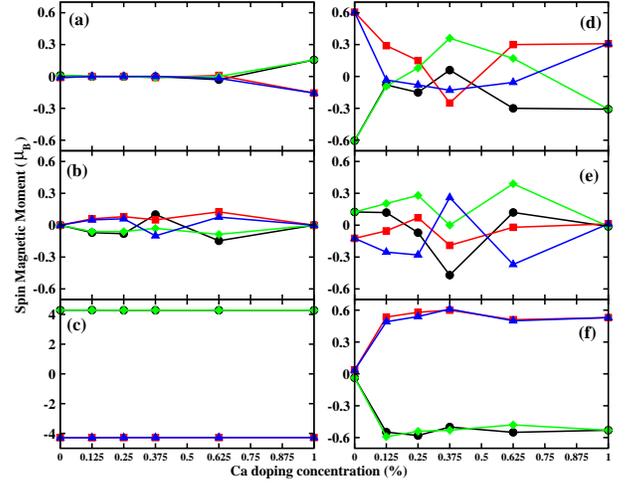}
	\caption{The spin magnetic moment calculated under GGA+$U$+SOC [001]  for AFM state, are shown for (a) Fe-$m_x$ (b) Fe-$m_y$ (c) Fe-$m_z$ (d) Ir-$m_x$ (e) Ir-$m_y$ (f) Ir-$m_z$ across the (Sr$_{1-x}$Ca$_x$)$_2$FeIrO$_6$ series.}
	\label{fig:Fig16}
\end{figure}

Since, the two end compounds i.e., SFIO ($x$ = 0) and CFIO ($x$ = 1) in present (Sr$_{1-x}$Ca$_x$)$_2$FeIrO$_6$ series show many similarities as well as dissimilarities, it would be interesting to explore the evolution of electronic structure across the series at microscopic level. Therefore, we have started with $\sqrt{2}$ $\times$ $\sqrt{2}$ $\times$ 1 supercell of SFIO, that contains eight Sr sites, four Fe and Ir sites each with total 40 atoms in the supercell. We have performed Ca-doping calculations for specific doping levels viz. 12.5\%, 25\%, 37.5\% and 62.5\%, due to constraint of the size of the supercell. The calculated GGA+$U$+SOC [spin quantization axis 001] DOS in AFM ground state are shown in Fig. \ref{fig:Fig15}. The electronic structure of the doped systems are not drastically different than the two end compounds i.e SFIO and CFIO. The Fig. \ref{fig:Fig15} clearly indicates that near Fermi energy the Ir-$5d$ states dominate similar to SFIO and CFIO. Also, as Ca doping concentration increases, the Ir-$t_{2g}$ conduction bands are pushed up and band gap increases which attains maximum value for the optimal level at $x$ = 0.375 doping. In addition, the bandwidth of the valance band shirks due to doping. This nonmonotonic evolution of band gap with $x$ explains the measured resistivity which shows the resistivity at low temperature maximizes around 40\% of Ca doping (Fig. \ref{fig:Fig11}).

\begin{figure}
	\centering
		\includegraphics[width=8.5cm]{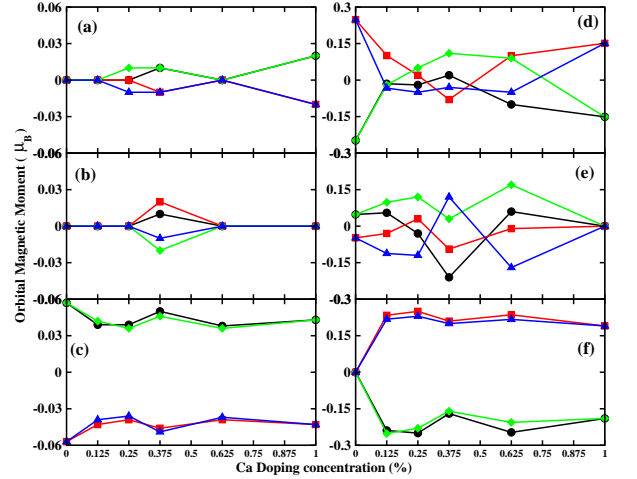}
	\caption{The orbital magnetic moment calculated under GGA+$U$+SOC [001] for AFM state, are shown for (a) Fe-$o_x$ (b) Fe-$o_y$ (c) Fe-$o_z$ (d) Ir-$o_x$ (e) Ir-$o_y$ (f) Ir-$o_z$ across the (Sr$_{1-x}$Ca$_x$)$_2$FeIrO$_6$ series.}
	\label{fig:Fig17}
\end{figure}

However, an interesting trend has been found in calculated spin and orbital magnetic moment as shown in Figs. \ref{fig:Fig16} and \ref{fig:Fig17}, respectively for whole series ($x$ = 0: SFIO; $x$ = 1: CFIO). From the Fig. \ref{fig:Fig16}c, it is very clear that the Fe spin magnetic moments are mainly aligned along $z$-component, while the $m_x$ and $m_y$ components (Figs. 16a and 16b) are very small compared to $m_z$ component. Also, it shows that in spite of Ca doping the $m_z$ component of Fe spin moment remains almost unchanged ($\sim$ 4.29 $\mu_B$/site) across the series. On the other side, Ir spin moments for $x$ = 0 (i.e., SFIO) are mainly concentrated in $m_x$ component with small contribution of $m_y$ component while almost there nothing comes in $m_z$ component. Therefore, for SFIO the Fe and Ir spins directions are orthogonal to each other. As the Ca concentration increases, the four Ir/Fe sites became inequivalent which results in an unequal development of moments at each site both in terms of magnitude and changing the sign. However, the effects are more prominent in the case of Ir than Fe due to its prominent SOC effect. For CFIO ($x$ = 1), the Ir spin moment are concentrated both in $m_x$ and $m_z$ components, lying in $xz$-plane. The orbital magnetic moment variation across the Ca doping has been shown in Fig. \ref{fig:Fig17}. Point to be mentioned here is that the orbital magnetic moments at Fe sites are very small ($\sim$ 0.06 $\mu_B$/site) compared to that of the Ir site ($\sim$ 0.24 $\mu_B$/site), which is expected due to the quenching of orbital degrees of freedom in Fe. However, the evolution of orbital magnetic moment with Ca doping follows very similar trend as that of the spin magnetic moment (see Fig. 16). Similar to the spin moment, the Ir orbital moment for SFIO also shows major contribution along the $m_x$ direction and for CFIO it lies in $m_x$ and $m_z$ directions as shown in Figs. \ref{fig:Fig17}d, \ref{fig:Fig17}e and \ref{fig:Fig17}f, respectively.

\begin{figure*}
	\centering
		\includegraphics[width=14cm]{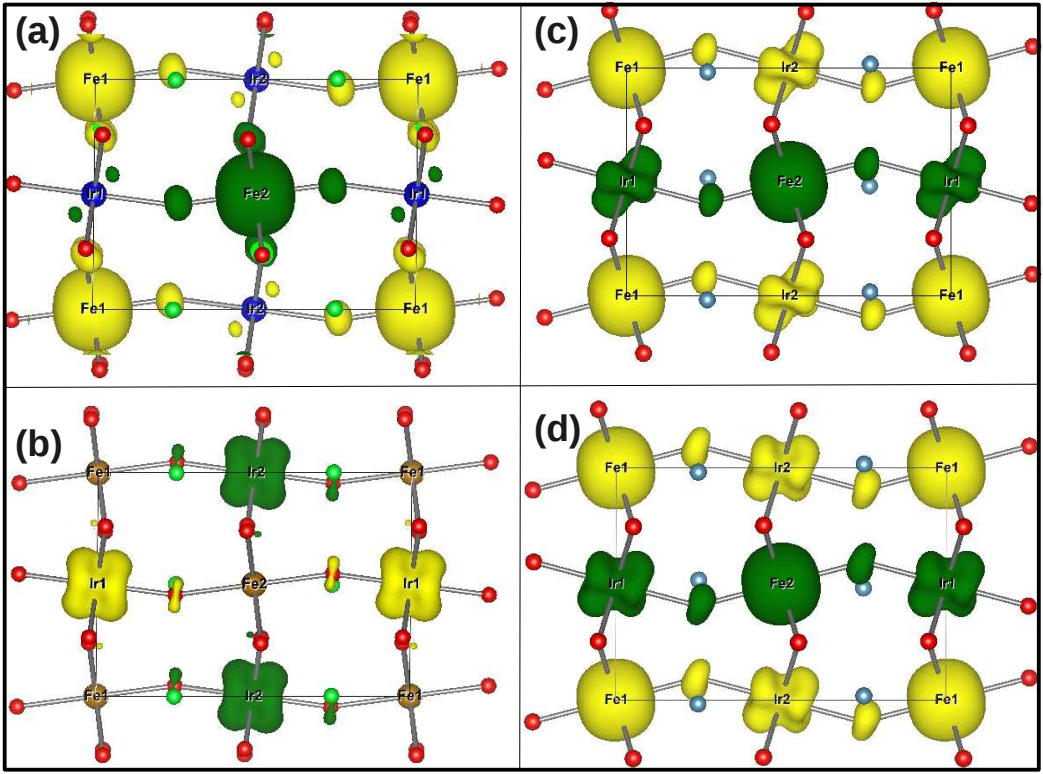}
	\caption{The calculated magnetization density plot for AFM state with GGA+$U$+SOC scheme. The energetically favored spin quantization axes are chosen for each compounds i.e SFIO: [100] and CFIO : [110].  (a) and (b) show the plot of $z$-component ($m_z$) and $x$-component ($m_x$) for SFIO, whereas (c) and (d) show similarly calculated $m_x$ and $m_y$, respectively for CFIO.}
	\label{fig:Fig18}
\end{figure*}

All the above GGA+$U$+SOC calculations with spin quantization axis along [001] direction, hint for a strong magnetocrystalline anisotropy in these materials. Therefore, we have done total energy comparison along different spin quantization axes for both SFIO and CFIO. Our results show that SFIO favors easy-axis ([001] and [100] are degenerate in energy) kind of magnetocrystalline anisotropy ($\sim$ 1.61 meV/f.u) whereas CFIO favors [110] easy-plane kind of magnetocrystalline anisotropy (2.52 meV/f.u). The calculated spin quantization results quite agree with the neutron diffraction measurements as shown in Fig.\ref{fig:Fig10}(c). In [001] spin quantized ground state configuration of SFIO, all the Fe moments (spin + orbital) point along the $z$-direction, however, the Ir moments (spin + orbital) remain orthogonal to that of Fe moment, directed mainly along $x$-direction with small component along $y$-direction. One point need to be cleared here, that if we consider [100], then also Fe and Ir moments develop in the orthogonal direction, but then Fe and Ir spin moments aligned in the $x$- and $z$-direction, respectively. On the other hand, in [110] spin quantized ground state of CFIO, the Fe and Ir spin moments are almost lying in $xy$-plane (Fe-$m_x$=3.04 $\mu_B$/site, Fe-$m_y$=3.02 $\mu_B$/site, Ir-$m_x$= 0.43 $\mu_B$/site, Ir-$m_y$=0.46 $\mu_B$/site), whereas the orbital moments for Ir are lying along the body the diagonal [111] plane (Ir-$o_x$= 0.15 $\mu_B$/site, Ir-$m_y$=0.14 $\mu_B$/site, Ir-$m_z$=0.17 $\mu_B$/site), as shown in Figs. \ref{fig:Fig18}(c,d).

These results can be visualized more clearly in plot of magnetization density, as presented in Fig. \ref{fig:Fig18}. Point to be remembered here is that, for plotting the Figs.\ref{fig:Fig18} we have chosen the lowest energy spin quantization axis for both SFIO (i.e [001] degenerate with [100]) and CFIO (i.e [110]) as discussed before, whereas in Fig. \ref{fig:Fig16} and Fig. \ref{fig:Fig17} the spin quantization axis is set to [001] for the whole series. For SFIO, Figs. \ref{fig:Fig18}a and \ref{fig:Fig18}b clearly show that Fe spin moments have only $z$-component, whereas Ir moments have nothing in $z$-component but completely concentrated in the $x$-component. The shape of the Fe isosurface is almost spherically symmetric as expected for the half-filled $d$ orbital ($d^5$) in $S$ = 5/2 spin state. However, the shape of the Ir isosurface is distorted. For CFIO, the magnetization density along the $m_x$ and $m_y$ are plotted in Figs. \ref{fig:Fig18}c and \ref{fig:Fig18}d, respectively. From the figure, it is evident that in case of CFIO, both Fe and Ir spin densities have almost similar contribution in both $m_x$ and $m_y$ component as discussed earlier, whereas the Fe and Ir spin moments are mainly concentrated in the $x$ and $z$ component. It is further observed that the shape of Fe isosurfaces are very similar in both cases of SFIO and CFIO, however, the shape of Ir isosurfaces vary significantly in both cases. The more distorted shape of the Ir isosurface in CFIO than SFIO, indicates that the
effective strength of SOC compared to crystal filed splitting and Coulomb correlation are substantially quenched in the case of CFIO due to enhanced octahedral distortion. Nonetheless, this evolution of magnetization density with Ca substitution or lattice distortion is quite intriguing. 

\section{Discussion and Conclusion}
In summary, solid state reaction method is used for the preparation of polycrystalline (Sr$_{1-x}$Ca$_x$)$_2$FeIrO$_6$ series with $x$ from 0 to 1. The sample quality has been checked with XRD measurements and Rietveld analysis which suggest triclinic-\textit{I$\bar{1}$} structural symmetry is retained for the whole series, though the lattice parameters evolve with composition. Analysis of XPS data and electronic structure calculations suggest Sr/Ca is in 2+ nominal charge state while Fe/Ir retains its 3+/5+ nominal charge state throughout the series. Magnetization measurements suggest AFM spin ordering remain to be low temperature magnetic state across the series though the transition temperature $T_N$ decreases and other magnetic parameters exhibit anomalous behavior around $x$ = 0.2. Temperature dependent neutron powder diffraction measurements reveal the triclinic-\textit{I$\bar{1}$} structural symmetry down to low temperature and confirm the magnetic Bragg peak at 5 K is related to AFM ordering where the site ordered moments for both Fe and Ir are observed which has also been reconfirmed by the electronic structure total energy calculations. All the samples are found to be Mott type insulator where the charge transfer mechanism follows 3D variable-range-hopping model. 

The GGA+$U$+SOC electronic structure calculations reveal detail microscopic mechanism of the AFM-insulating state. We find that for present investigated materials, there are strong competition among the SOC, electronic correlation ($U$), structural distortion and band width. As a result, the effective strength of SOC is reduced and it is not the most dominating energy scale of the systems compared to other energy scales involved here. In conventional SOC-driven insulators i.e., materials containing Ir$^{4+}$ (5$d^5$) state,\cite{Kim,Kim1,Krempa,Rau,Yogesh,Pesin} the orbital magnetic moment is as high as almost two times than the spin magnetic moment due to very high SOC. In the present systems, however, we find the orbital magnetic moment is lower than the spin magnetic moment (almost half), which indicates a considerably weaker strength of the effective SOC. Although, the strength of SOC is an inherent atomic property, which does not change for a particular element in a particular electronic configuration, but the comparative strength of the other energy scales such as $U$, band width and structural distortions dictate relative weightage in the final outcome. The present series of compounds are highly distorted and having large band width ($\sim$ 1 eV) compared to the conventional SOC-driven insulating states found in other iridates (i.e., Sr$_2$IrO$_4$, A$_2$Ir$_2$O$_7$, Na$_2$IrO$_3$, etc.)\cite{Kim,Kim1,Krempa,Rau,Yogesh,Pesin}. Additionally, these double perovskites contain both 3$d$ (Fe) and 5$d$ (Ir) transition metals, hence an interplay between the electronic correlation and SOC as well as the narrow and extended wave functions plays vital role in deciding their magnetic and electronic properties\cite{Krempa,Rau}. However, the SOC is an important energy scale here too, as it is very crucial in stabilizing the non-collinear AFM ground state with substantial anisotropy found in the calculations, but not the main energy scale of the system. Therefore, these materials belong to the intermediate region between conventional electronic-correlation-driven Mott-insulator and conventional SOC-driven Mott-insulator and can be classified as the `SOC enhanced correlation-driven AFM-Mott Insulator'. 

Our calculations further show SOC induced spin magnetic moment on Ir$^{5+}$ ($\sim$ 0.62 $\mu_B$/site) also agrees with the NPD experiments showing a site ordered moment of Ir ($\sim$ 0.5 $\mu_B$). An indirect band gap between $\Gamma$ and $M$ point has been captured in band calculations which further increases with Ca doping. The calculations under GGA+$U$+SOC+AFM scheme show an interesting evolution of magnetic moment components ($m_x$, $m_y$ and $m_z$) of both Fe and Ir in present series which mainly arises due to local structural distortion caused by lower size Ca$^{2+}$ substitution. The calculated magnetization density further shows the deviation from ideal $J_{eff}$ quantum number description of the electronic state due to the strong interplay among the SOC, structural distortion and electron correlation in present series of materials. 

\section{Acknowledgment}
We acknowledge FIST-DST and PURSE-DST program for supporting `Low temperature and high magnetic field facility' and `Small-scale helium liquefier facility', respectively. We thank Dr. Alok Banerjee, UGC-DAE CSR, Indore and IIT, Delhi for the magnetization measurements. We are thankful to Dhruva reactor, Trombay, India for the neutron measurements and UGC-DAE CSR, Mumbai for the financial support under the Collaborative Research Scheme (No. CRS-M-268). KCK and HK acknowledge UGC, India for financial support. RR acknowledges IIT Goa (MHRD, Govt. of India) for providing the fellowship. Author SK thanks Department of Science and Technology (DST), Govt. of India for providing INSPIRE research funding (Grant No. DST/INSPIRE/04/2016/000431; IFA16-MS91).


\begin{thebibliography}{}
\bibitem{Longo} J. Longo, and R. Ward, J. Am. Chem. Soc. \textbf{83},2816 (1961).
\bibitem{Sleight} A. W. Sleight, and J. F. Weiher, J. Phys. Chem. Solids \textbf{33},679 (1972).
\bibitem{Yokoyama} H. Yokoyama, and T. Nakagawa, J. Phys. Soc. Japan \textbf{28},1197 (1970).
\bibitem{Serrate} D. Serrate, J. M. D. Teresa and M. R. Ibarra, J. Phys. Condens. Matter \textbf{19},2 (2016).
\bibitem{Vasala} S. Vasala, M. Karppinen, Prog. Solid State Chem. \textbf{43} 1 (2014).
\bibitem{Gopal} J. Gopalakrishnan, A. Chattopadhyay, S. B. Ogale, T. Venkatesan, R. L. Greene, A. J. Millis, K. Ramesha, B. Hannoyer and G. Marest, Phys. Rev. B \textbf{62}, 9538 (2000).
\bibitem{Battle} P. D. Battle, T. C. Gibb, C. W. Jones, and F. Studer, J. Solid State Chem. \textbf{78} 281 (1989).
\bibitem{Nomura} K. Nomura, R. Zboril, J. Tucek, W. Kosaka, S. Ohkoshi, and I.Felner, J. Appl. Phys. \textbf{102}, 013907 (2007).
\bibitem{Naveen} K. Naveen, M. Reehuis, P. Adler, P. Pattison, A. Hoser, T. K. Mandal, U. Arjun, P. K. Mukharjee, R. Nath, C. Felser, and A. K. Paul, Phys. Rev. B \textbf{98}, 224423 (2018).
\bibitem{Paul} A. K. Paul, M. Reehuis, V. Ksenofontov, B. Yan, A. Hoser, D. M. T$\ddot{o}$bbens, P. M. Abdala, P. Adler, M. Jansen, and C. Felser, Phys. Rev. Lett. \textbf{111}, 167205 (2013).
\bibitem{Feng} H. L. Feng, M. Arai, Y. Matsushita, Y. Tsujimoto, Y. Guo, C. I. Sathish, X. Wang, Y. H. Yuan, M. Tanaka, and K. Yamaura, J. Am. Chem. Soc. \textbf{136},3326 (2014).
\bibitem{Kanungo1} S. Kanungo, B. Yan, M. Jansen, and C. Felser, Phys. Rev. B \textbf{89}, 214414 (2014).
\bibitem{Cao1} G. Cao, S. McCall, M. Shepard, J. E. Crow and R. P. Guertin, Phys. Rev. B \textbf{56}, 321 (1997).
\bibitem{Gat} I. M. Gat-Malureanu, J. P. Carlo, T. Goko, A. Fukaya, T. Ito, P. P. Kyriakou, M. I. Larkin, G. M. Luke, P. L. Russo, A. T. Savici, C. R. Wiebe, K. Yoshimura, and Y. J. Uemura, Phys. Rev. B \textbf{84}, 224415 (2011).
\bibitem{Fuchs} D. Fuchs, M. Wissinger, J. Schmalian, C.-L. Huang, R. Fromknecht, R. Schneider, and H. v. L$\ddot{o}$hneysen, Phys. Rev. B \textbf{89}, 174405 (2014). 
\bibitem{Terasaki} I. Terasaki, S. Shibasaki, S. Yoshida, and W. Kobayashi, Materials \textbf{3}, 786 (2010).
\bibitem{Ma} C. Ma, H. X. Yang, L. J. Zeng, Y. Zhang, L. L. Wang, L. Chen, R. Xiong, J. Shi, and J. Q. Li,J. Phys. Condens. Matter \textbf{21},215606 (2009).
\bibitem{Deng} G. Deng, N. Tsyrulin, P. Bourges, D. Lamago, H. Ronnow, M. Kenzelmann, S. Danilkin, E. Pomjakushina, and K. Conder, Phys. Rev. B \textbf{88}, 014504 (2013).
\bibitem{Liu} Y. Liu, Z. Yang, H. Yang, T. Zou, Y. Xie, B. Chen, Y. Sun, Q. Zhan, and R. Li, J. Phys. D: Appl. Phys. \textbf{45}, 245001 (2012).
\bibitem{Jia} S. Jia, A. J. Williams, P. W. Stephens, and R. J. Cava, Phys. Rev. B \textbf{80}, 165107 (2009).
\bibitem{Philipp} J. B. Philipp, P. Majewski, L. Alff, A. Erb, R. Gross, T. Graf, M. S. Brandt, J. Simon, T. Walther, W. Mader, D. Topwal,and D. D. Sarma, Phys. Rev. B \textbf{68}, 144431 (2003).
\bibitem{Taylor} A.E.Taylor, R. Morrow, R. S. Fishman, S. Calder, A. I. Kolesnikov, M. D. Lumsden, P. M. Woodward and A. D. Christianson, Phys. Rev. B \textbf{93}, 220408(R) (2016).
\bibitem{Chen} G. Chen, R. Pereira, L. Balents, Phys. Rev. B \textbf{82} 174440 (2010).
\bibitem{Krock} Y. Krockenberger, K. Mogare, M. Reehuis, M. Tovar, M. Jansen, G. Vaitheeswaran, V. Kanchana, F. Bultmark, A. Delin, F. Wilhelm, A. Rogalev, A. Winkler, and L. Alff, Phys. Rev. B \textbf{75}, 020404 (2007).
\bibitem{Krempa} W. Witczak-Krempa, G. Chen, Y. B. Kim and L. Balents, Annu. Rev. Condens. Matter Phys. \textbf{5}, 57 (2014)
\bibitem{Rau} J. G. Rau, E. Kin-Ho Lee, and H.-Y. Kee, Annu. Rev. Condens. Matter Phys. \textbf{7}, 195 (2016)
\bibitem{Kim} B. J. Kim, Hosub Jin, S. J. Moon,J.Y. Kim,B.-G. Park,C. S. Leem, Jaejun Yu,T.W. Noh,C. Kim,S.J. Oh,J.H.Park,V. Durairaj,G. Cao, and E. Rotenberg, Phys. Rev. Lett. \textbf{101}, 076402 (2008).
\bibitem{Kim1} B. J. Kim, H. Ohsumi, T. Komesu, S. Sakai, T. Morita, H.Takagi, T. Arima, Science \textbf{323}, 1329 (2009).
\bibitem{Kharkwal} K. C. Kharkwal, and A. K. Pramanik, J. Phys. Condens. Matter \textbf{29}, 495801, (2017).
\bibitem{Wallace} D. Wallace and T. M. McQueen, Dalton Trans. 47, 20344 (2015).
\bibitem{Nag} A. Nag, S. Middey, S. Bhowal, S. K. Panda, R. Mathieu, J. C. Orain, F. Bert, P. Mendels, P. G. Freeman, M. Mansson, H. M. Ronnow, M. Telling, P. K. Biswas, D. Sheptyakov, S. D. Kaushik, V. Siruguri, C. Meneghini, D. D. Sarma, I. Dasgupta and S. Roy, Phys. Rev. Lett. \textbf{116}, 097205 (2016). 
\bibitem{Khaliullin} G. Khaliullin, Phys. Rev. Lett. \textbf{113}, 197201 (2013).
\bibitem{Cao} G. Cao, T. F. Qi, L. Li, J. Terzic, S. J. Yuan, L. E. DeLong, G. Murthy, and R. K. Kaul, Phys. Rev. Lett. \textbf{112}, 056402 (2014).
\bibitem{Ranjbar} B. Ranjbar, E. Reynolds, P. Kayser, and B. J. Kennedy, Inorg. Chem. \textbf{54}, 10468 (2015).
\bibitem{Dey} T. Dey, A. Maljuk, D. V. Efremov, O. Kataeva, S. Gass, C. G. F. Blum, F. Steckel, D. Gruner, T. Ritschel,
A. U. B. Wolter, J. Geck, C. Hess, K. Koepernik, J. van den Brink, S. Wurmehl, and B. B$\ddot{u}$chner, Phys. Rev. B \textbf{93}, 014434 (2016).
\bibitem{Phelan} B. F. Phelan, E. M. Seibel, D. B. Jr., W. Xie, and R. J. Cava, Solid State Comm. \textbf{236},37 (2016).
\bibitem{Kanungo2} S. Kanungo, K. Mogare, B. Yan, M. Reehuis, A. Hoser, C. Felser, and M. Jansen, Phys. Rev. B \textbf{93}, 245148 (2016).
\bibitem{Sala} M. M. Sala, K. Ohgnshi, A. Al-Xein, Y. Hirata, G. Monaco and M. Krisch, Phys. Rev. Lett. \textbf{112}, 176402 (2014). 
\bibitem{Kusch} M. Kusch, V. M. Katukuri, N. A. Bogdanov, B. B$\ddot{u}$chner, T. Dey, D. V. Efremov, J. E. H. Borrero, B. H. Kim, M. Krisch, A. Maljuk, M. Moretti Sala, S. Wurmehl, G. A. Cansever, M. Sturza, L. Hozoi, J. van den Brink, and J. Geck, Phys. Rev. B \textbf{97}, 064421 (2018).
\bibitem{Kim2} B. H. Kim, D. V. Efremov, and J. van den Brink, Phys. Rev. Materials \textbf{3}, 014414 (2019).
\bibitem{Shannon} R. D. Shannon, Acta Cryst. \textbf{A32}, 751 (1976).
\bibitem{Battle1} P. D. Battle, G. R. Blake, T. C. Gibb  and J. F. Vente, J. Solid state Chem. \textbf{145}, 541 (1999).
\bibitem{Qasim} I. Qasim, P. E. R. Blanchard, S. Liu, C. Tang, B. J. Kennedy, M. Avdeev, J. A. Kimpton 2013 J.solid state Chem. \textbf{206}, 242 (2013).
\bibitem{Bufaical} L. Bufaiçal, C. Adriano, R. L. Serrano, J. G. S. Duque, L. M. Ferreira, C. R. Ayala, E. B. Saitovitch, E. M. Bittar and P. G. Pagliuso, J. Solid State Chem. \textbf{212}, 23 (2014). 
\bibitem{Kayser} P. Kayser, J. A. Alonso, F. J. Mompean, M. Retuerto, M. Croft, A. Ignatov, and M. T. F. Diaz, Eur. J. Inorg. Chem.  \textbf{2015}, 5027 (2015).
\bibitem{Laghuna} M. A. L. Marco, P. Kayser, J. A. Alonso, M. A. M. Lope, M. V. Veenendaal, Y. Choi and D. Haskel, Phys. Rev. B \textbf{91} 214433(2015). 
\bibitem{Avijit} A. K. Paul, M. Jansen, B. Yan, C. Felser, M. Reehuis and P. M. Abdala, Inorg. Chem. \textbf{52}, 6713 (2013).
\bibitem{Morrow} R. Morrow, R. Mishra, O. D. Restrepo, M. R. Ball, W. Windl, S. Wurmehl, U. Stockert, B. B$\ddot{u}$chner and P. M. Woodward, J. Am. Chem. Soc. \textbf{135}, 18824 (2013).
\bibitem{vasp} G. Kresse and J. Hafner; Phys. Rev. B {\bf47}, R558 (1993); G. Kresse and J. Furthmuller ibid. {\bf54}, 11169 (1996).
\bibitem{PBE} J. P. Perdew, K. Burke and M. Ernzerhof; Phys. Rev. Lett. {\bf77}, 3865 (1996).
\bibitem{GGAU2} S. L. Dudarev, G. A. Botton, S. Y. Savrasov, C. J. Humphreys and A. P. Sutton, Phys. Rev. B {\bf57}, 1505 (1998).
\bibitem{GGAU1} V. I. Anisimov, I. V. Solovyev, M. A. Korotin, M. T. Czyzyk and G. A. Sawatzky, Phys. Rev. B {\bf48}, 16929 (1993).
\bibitem{Celorrio} V. Celorrio, L. Calvillo, G. Granozzi,  A. E. Russell, and  D. J. Fermin, Topics in Catalysis \textbf{61}, 154 (2018).
\bibitem{Doveren} H. V. Doveren, and J. A. T. Verhoeven, J. Elec. Spec. and Related Phenom., \textbf{21}, 265 (1980). 
\bibitem{Liu1} X. Liu, Yanwei Cao, B. Pal, S. Middey, M. Kareev, Y. Choi, P. Shafer, D. Haskel, E. Arenholz, and J. Chakhalian, Phys. Rev. Materials \textbf{1}, 075004 (2017).
\bibitem{Rogge} P. C. Rogge, R. U. Chandrasena, A. Cammarata, R. J. Green, P. Shafer, B. M. Lefler, A. Huon, A. Arab, E. Arenholz, H. N. Lee, T-L. Lee, S. Nem$\check{s}\acute{a}$k, J. M. Rondinelli, A. X. Gray, and S. J. May, Phys. Rev. Materials \textbf{2}, 015002 (2018).
\bibitem{Ghaffari} M. Ghaffari, H. Huang, O. K. Tanaand and M. Shannon, CrystEngComm, \textbf{14}, 7487 (2012).
\bibitem{Tsuyama} T. Tsuyama, T. Matsuda, S. Chakraverty, J. Okamoto, E. Ikenaga, A. Tanaka, T. Mizokawa, H. Y. Hwang, Y. Tokura and H. Wadati, Phys. Rev. B \textbf{91}, 115101 (2015).
\bibitem{Yamashita} T. Yamashita, and P. Hayes, App. Surf. Science, \textbf{254}, 2441 (2008). 
\bibitem{Grosvenor} A. P. Grosvenor, B. A. Kobe, M. C. Biesinger and N. S. McIntyre, Surf. Interface Anal., \textbf{36}, 1564 (2004).
\bibitem {Mullet} M. Mullet, V. Khare and C. Ruby, Surf. Inter. Anal., \textbf{40}, 323 (2008).
\bibitem{Biesingera} M. C. Biesingera, B. P. Paynec, A. P. Grosvenord, L. W. M. Laua, A. R. Gersonb, R. S. C. Smartb, App. Surf. Sciences, \textbf{257},2717 (2011).
\bibitem{Fuentes} S. Fuentes, P. Munoz, N. Barraza, E. C. Angel, and C. M. S. Torres,J. Sol-Gel Sci. Technol. \textbf{75},593 (2015).
\bibitem{Zhu} W. K. Zhu, M. Wang, B. Seradjeh, F. Yang, and S. X. Zhang, Phys. Rev. B \textbf{90}, 054419 (2014).
\bibitem{Imtiaz} I. N. Bhatti, R. S. Dhaka and A. K. Pramanik, Phys. Rev. B \textbf{96}, 144433 (2017).
\bibitem{Harish} H. Kumar, R. S. Dhaka and A. K. Pramanik, Phys. Rev. B \textbf{95}, 054415 (2017)  
\bibitem{Otsubo} T. Otsubo, S. Takase, and Y. Shimizu, ECS Transaction, \textbf{3}, 263 (2006).
\bibitem{Cao2} G. Cao, V. Durairaj, S. Chikara, L. E. DeLong, S. Parkin, and P. Schlottmann, Phys. Rev. B \textbf{76}, 100402 (2007).
\bibitem{Arrott} A. Arrott, Phys. Rev. \textbf{108}, 1394 (1957).
\bibitem{Kayser1} P. Kayser, M. J. Mart$\acute{i}$nez-Lope, J. A. Alonso, M. Retuerto, M. Croft, A. Ignatov, and M. T. Fernandez-D$\acute{i}$az, Inorg. Chem. \textbf{52}, 11013 (2013). 
\bibitem{chen1} J. Chen, H. J. Qin, F. Yang, J. Liu, T. Guan, F. M. Qu, G. H. Zhang, J. R. Shi, X. C. Xie, C. L. Yang, K. H. Wu, Y. Q. Li, and L. Lu, Phys. Rev. Lett. \textbf{105}, 176602 (2010).
\bibitem{he} Hong-Tao He, Gan Wang, Tao Zhang, Iam-Keong Sou, George K. L Wong, Jian-Nong Wang, Hai-Zhou Lu, Shun-Qing Shen, and Fu-Chun Zhang, Phys. Rev. Lett. \textbf{106}, 166805 (2011).
\bibitem{jenderka} M. Jenderka, J. Barzola-Quiquia, Z. Zhang, H. Frenzel, M. Grundmann, and M. Lorenz, Phys. Rev. B \textbf{88}, 045111 (2013).
\bibitem{nguyen} V. L. Nguyen, B. Z. Spivak and B. I. Shklovskii, Zh. Eksp. Teor. Fiz. \textbf{89}, 1770 (1985 )
\bibitem{Mott} N. Mott,  Conduction in non-crystalline materials, Clarendon press,Oxford (1993). 
\bibitem{Wang} G. J. Wang, C. C. Wang, S. G. Huang, X. H. Sun, C. M. Lei, T. Li, and J. Y. Mei, J. Electroceram.\textbf{28},172(2012).
\bibitem{Lin} Y. Q. Lin and X. M. Chen, J. Am. Ceram. Soc. \textbf{92}, 782 (2011).
\bibitem{Full} FULLPROF suite, http://www.ill.eu/sites/fullprof/.
\bibitem{Yogesh} Y. Singh and P. Gegenwart, Phys. Rev. B \textbf{82}, 064412 (2010)
\bibitem{Pesin} D. Pesin and L. Balents, Nat. Phys. \textbf{6}, 376 (2010) 
\bibitem{Imada} M. Imada, A. Fujimori and Y. Tokura, Rev. Mod. Phys. \textbf{70}, 1039 (1998).
\bibitem{Narayanan} N. Narayanan, D. Mikhailova, A. Senyshyn, D. M. Trots, R. Laskowski, P. Blaha, K. Schwarz, H. Fuess, and H. Ehrenberg, Phys. Rev B \textbf{82}, 024403 (2010).
\bibitem{GCao} G. Cao, A. Subedi, S. Calder, J.-Q. Yan, J. Yi, Z. Gai, L. Poudel, D. J. Singh, M. D. Lumsden, A. D. Christianson,B. C. Sales, and D. Mandrus, Phys. Rev. B \textbf{87}, 155136 (2013).
\bibitem{WKZhu} W. K. Zhu, J.-C. Tung, W. Tong, L. Ling, M. Starr, J. M.Wang, W. C. Yang, Y. Losovyj, H. D. Zhou, Y. Q. Wang, P.-H. Lee, Y.-K. Wang, Chi-Ken Lu, and S. X. Zhang, arXiv:1608.07763 (2016).
\bibitem{goodenough} J. B. Goodenough, Phys. Rev \textbf{100}, 564 (1955).
\bibitem{kanamori} J. Kanamori, J. Phys. Chem. Solids \textbf{10}, 87 (1959).
\end{thebibliography}
\end{document}